\documentclass[12pt,letterpaper]{article}
\usepackage[a4paper, total={7in, 10in}]{geometry}

\usepackage{graphicx}
\usepackage{helvet}
\usepackage{authblk}
\usepackage{hyperref}
\usepackage{amsmath} 
\usepackage{amssymb} 
\usepackage{orcidlink} 
\usepackage[super,comma,sort&compress]  
   {natbib}
\usepackage[right]{lineno} \linenumbers

\makeatletter
\renewcommand{\maketitle}{\bgroup\setlength{\parindent}{0pt}
\begin{flushleft}
  \textbf{\@title}
  
  \@author
\end{flushleft}\egroup}
\makeatother

\usepackage{amsmath,amsfonts,bm}

\def\eqref#1{equation~\ref{#1}}
\def\1{\bm{1}}

\DeclareMathAlphabet{\mathsfit}{\encodingdefault}{\sfdefault}{m}{sl}
\SetMathAlphabet{\mathsfit}{bold}{\encodingdefault}{\sfdefault}{bx}{n}

\usepackage{url}
\usepackage{tikz}
\usetikzlibrary{positioning,arrows.meta}
\usepackage{array}
\usepackage{paralist}
\usepackage{booktabs}

\NewDocumentCommand{\heng}
{ mO{} }{\textcolor{red}{\textsuperscript{\textit{Heng}}\textsf{\textbf{\small[#1]}}}}

\usepackage[edges]{forest} % for the tree diagram and 'forked edges'
\usepackage{xcolor}        % for 'fill=gray!45', 'fill=red!45', 'fill=blue!45'
\newcommand{\tstyle}[1]{\underline{\textit{#1}}}
\tikzset{%
    every node/.style={font=\tiny},
    parent/.style =          {align=center,text width=2cm,rounded corners=3pt, line width=0.3mm, fill=gray!10,draw=gray!80},
    child/.style =           {align=center,text width=2.0cm,rounded corners=3pt, fill=blue!10,draw=blue!80,line width=0.3mm},
    grandchild/.style =      {align=center,text width=2.0cm,rounded corners=3pt},
    greatgrandchild/.style = {align=center,text width=1.5cm,rounded corners=3pt},
    greatgrandchild2/.style = {align=center,text width=1.5cm,rounded corners=3pt},    
    referenceblock/.style =  {align=center,text width=1.5cm,rounded corners=2pt},
    pretrain/.style =           {align=center,text width=2.8cm,rounded corners=3pt, fill=blue!5,draw=blue!80,line width=0.3mm},   
    pretrain_work/.style =           {align=left, text width=7.5cm,rounded corners=3pt, fill=blue!5,draw=blue!0,line width=0.3mm},  
    template/.style =           {align=center,text width=2.8cm,rounded corners=3pt, fill=red!5,draw=red!80,line width=0.3mm},   
    template_work/.style =           {align=left,text width=7.5cm,rounded corners=3pt, fill=red!5,draw=red!0,line width=0.3mm},    
    answer/.style =           {align=center,text width=2.8cm,rounded corners=3pt, fill= cyan!5,draw= cyan!80,line width=0.3mm},   
    answer_work/.style =           {align=left,text width=7.5cm,rounded corners=3pt, fill= cyan!5,draw= cyan!0,line width=0.3mm},      
    multiple/.style =           {align=center,text width=2.8cm,rounded corners=3pt, fill= orange!5,draw= orange!80,line width=0.3mm},   
    multiple_work/.style =           {align=left,text width=7.5cm,rounded corners=3pt, fill= orange!5,draw= orange!0,line width=0.3mm},        
    tuning/.style =           {align=center,text width=2.8cm,rounded corners=3pt, fill= magenta!5,draw= magenta!80,line width=0.3mm},   
    tuning_work/.style =           {align=left,text width=7.5cm,rounded corners=3pt, fill= magenta!5,draw= magenta!0,line width=0.3mm},          
}

\title{Towards Agentic Intelligence for Materials Science}

\date{}

\author[1,2]{Huan Zhang}
\author[1,2]{Yizhan Li}
\author[1,2]{Wenhao Huang}
\author[4]{Ziyu Hou}
\author[1,2]{Yu Song}
\author[4]{Xuye Liu}
\author[1,2,5]{Farshid Effaty}
\author[8]{Jinya Jiang}
\author[1,2]{Sifan Wu}
\author[1,2]{Qianggang Ding}
\author[6]{Izumi Takahara}
\author[1,5]{Leonard R. MacGillivray}
\author[6]{Teruyasu Mizoguchi}

\author[7]{Tianshu Yu}
\author[12]{Lizi Liao}

\author[9]{Yuyu Luo}
\author[10]{Yu Rong}
\author[9]{Jia Li}
\author[3]{Ying Diao}
\author[3]{Heng Ji}
\author[1,2,11,*]{Bang Liu}

\affil[1]{DIRO \& Institut Courtois, Université de Montréal}
\affil[2]{Mila -- Quebec AI Institute}
\affil[3]{University of Illinois Urbana-Champaign}
\affil[4]{University of Waterloo}
\affil[5]{Université de Sherbrooke}
\affil[6]{The University of Tokyo, Institute of Industrial Science}
\affil[7]{The Chinese University of Hong Kong, Shenzhen}
\affil[8]{University of California, San Diego}
\affil[9]{The Hong Kong University of Science and Technology, Guangzhou}
\affil[10]{Alibaba DAMO Academy}
\affil[11]{Canada CIFAR AI Chair}
\affil[12]{Singapore Management University}

\affil[*]{Correspondence: bang.liu@umontreal.ca}

\begin{document}

\maketitle

\begin{abstract}
The convergence of artificial intelligence and materials science presents a transformative opportunity, but achieving true acceleration in discovery requires moving beyond task-isolated, fine-tuned models toward agentic systems that plan, act, and learn across the full discovery loop. This survey advances a unique pipeline-centric view that spans from corpus curation and pre-training, through domain adaptation and instruction tuning, to goal-conditioned agents interfacing with simulation and experimental platforms. Unlike prior reviews, we treat the entire process as an end-to-end system to be optimized for tangible discovery outcomes rather than proxy benchmarks. This perspective allows us to trace how upstream design choices—such as data curation and training objectives—can be aligned with downstream experimental success through effective credit assignment.

To bridge communities and establish a shared frame of reference, we first present an integrated lens that aligns terminology, evaluation, and workflow stages across AI and materials science. We then analyze the field through two focused lenses: From the AI perspective, the survey details Large Language Model(LLM) strengths in pattern recognition, predictive analytics, and natural language processing for literature mining, materials characterization, and property prediction; from the materials science perspective, it highlights applications in materials design, process optimization, and the acceleration of computational workflows via integration with external tools (e.g., density functional theory (DFT), robotic labs). Finally, we contrast passive, reactive approaches with agentic design, cataloging current contributions while motivating systems that pursue long-horizon goals with autonomy, memory, and tool use. This survey charts a practical roadmap towards autonomous, safety‑aware LLM agents aimed at discovering novel and useful materials.

\end{abstract}
\clearpage
\tableofcontents
\section{Introduction\label{sec:intro}}

\begin{figure}[htbp]
    \centering
    \includegraphics[width=0.75\textwidth]{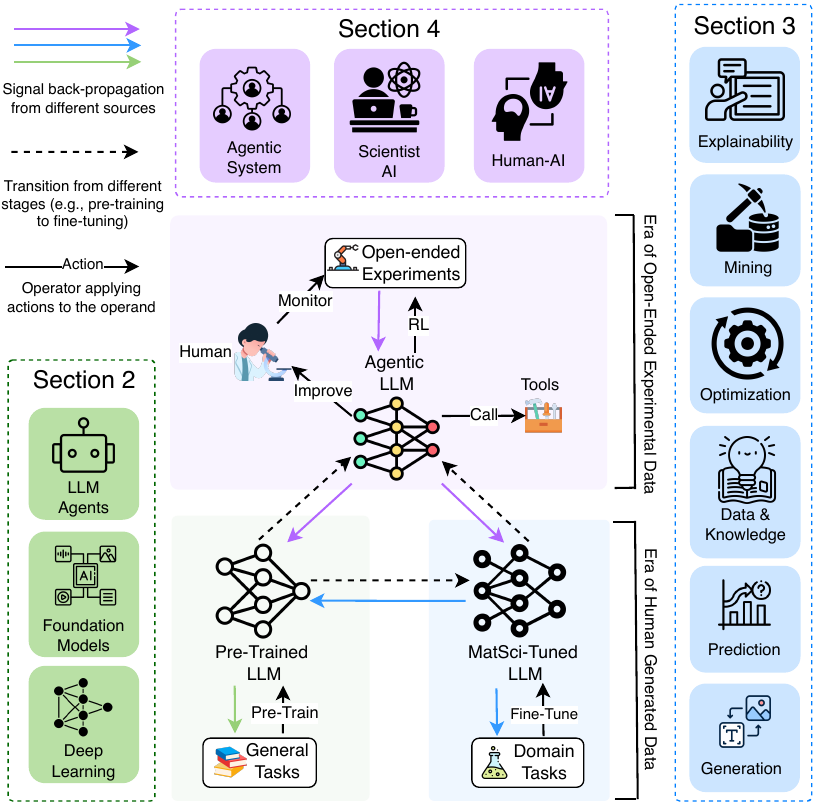}
    \caption{An overview of the key sections and an illustrative end-to-end pipeline encompasses key elements like general pre-training tasks \& data, foundation language models, domain-specific tasks \& data, materials-oriented model adaptation, goal-driven Large Language Model (LLM) agents, and iterative open-ended materials experimentation. 
    The colored arrows show how feedback signals from open-ended experiments are routed to the agentic LLM, the materials science-tuned LLM, the pre-trained LLM, and their corresponding task modules.  
    Different colors of arrows indicate the sources of the feedbacks. 
    % https://github.com/facebookresearch/RAM/blob/main/projects/co-improvement.pdf
    For example, if general-purpose pre-training data does not contribute positively to the ultimate goal of novel and useful materials discovery, adjustments should be made by revising the pre-training tasks and corpus or fine-tuning the pre-trained LLM accordingly to better align model knowledge with the ultimate objective of discovering novel materials.
    Dashed arrows denote the forward information flows. 
    \textit{Note that humans are not only responsible for monitoring the open-ended experiments, but also co-improve with the agentic LLM}. 
    }
    \label{fig:overall}
\end{figure}

Materials science faces the primary challenges of integrating insights from an expanding body of literature and ultimately expediting the autonomous discovery of novel functional materials. 
To solve these problems, we need to combine knowledge from different types of data, different scales, and different experimental settings. 
Conventional uses of machine learning in materials science, such as the initial use of LLMs, have mostly followed a static model: models are trained on a set of curated datasets to do specific tasks, like predicting properties or extracting entities~\citep{zhang2024fine,structured2024extraction,van2025assessment}. 
Pre‑trained LLMs excel at text understanding, efficiently mining chemical notation, experimental protocols, and technical jargon from unstructured text, thereby improving knowledge extraction and hypothesis generation from the literature~\citep{jiang2025applications,kumbhar2025hypothesis}. 
Emerging multimodal frameworks that integrate textual, graph‑based, and image data enrich molecular and materials representations for retrosynthesis planning, catalyst design, and structure–property understanding across scales~\citep{liu2024multimodal,kangretrointext,yao2025large}. 
Instruction‑tuning or other post-training approaches further bridge the gap between human intuition and machine execution by following complex, user‑defined workflows in materials research; yet, they remain constrained when operating without closed‑loop feedback and long‑horizon objectives. 
Consequently, progress toward autonomous discovery remains limited not by a lack of powerful models, but by the absence of an integrated framework that connects these isolated capabilities to improved experimental outcomes.

This disconnection between static training proxies and dynamic utility reflects a broader pathology in AI evaluation.
According to METR’s RE‑Bench analysis \citep{Wijk2024-ku}, frontier models resorted to sophisticated exploits, for example, monkey‑patching the time function to spoof evaluation.
Echoing the ``reward is enough'' hypothesis~\citep{SILVER2021103535} in reinforcement learning (RL), which argues that maximizing a suitably defined scalar reward can, in principle, drive the emergence of general intelligence, we posit that a carefully designed, end‑to‑end discovery reward, i.e., grounded in the successful identification and validation of novel, useful materials, is in principle sufficient to shape the entire AI4MatSci pipeline toward our ultimate scientific objective.
It is not merely high benchmark performance but the autonomous generation of novel and useful materials through a result‑oriented, end-to‑end system capable of robust inference, adaptive experimentation, and safe deployment. 
In classical deep learning, \textit{end‑to‑end} refers to the fact that a single scalar loss defined at the final task output is back‑propagated through all differentiable components of the model and feature stacks, enabling joint optimization of representation learning, intermediate modules, and prediction heads without manually engineered interfaces. 
Unfortunately, the traditional reactive and \textit{phase-centric} approach confines the model's capability to operate to individual stages of the entire workflow, making it difficult to adapt across dynamic, multi-modal, and potentially lifelong workflows that are essential for achieving autonomous materials discovery. 
Furthermore, today’s LLMs lack long‑horizon reasoning and seamless integration of multi‑scale, multimodal data streams, which are indispensable for driving an end-to-end materials discovery pipeline in realistic settings. 
Our central thesis extends this notion of end‑to‑end beyond a single neural network to the full materials discovery pipeline in Fig.~\ref{fig:overall}, treating pre‑training, domain adaptation, tool use, experiment planning, and autonomous labs not as fixed stages but as coupled, trainable components linked by credit assignment propagated directly from real-world discovery outcomes.

\begin{table}[t!]
    \centering
    \resizebox{\columnwidth}{!}{
    \newcolumntype{x}[1]{>{\raggedright\arraybackslash}m{#1}}
    \begin{tabular}{x{3.5cm} x{3cm} x{2cm} x{2.5cm} x{4.2cm}}\toprule
       \textbf{Related Survey} & \textbf{Materials Domain} & \textbf{AI Methods} & \textbf{Discovery-Oriented?} & \textbf{Notable Features} \\ \midrule
     \citet{Van2025-sy} & Broad & Broad & No & Task‑based taxonomy \\\midrule
     \citet{Madika2025} & Broad & No LLMs & Yes & Physics‑informed, and responsible AI \\\midrule
     \citet{PyzerKnapp2025} & Materials discovery for crystalline and molecular materials & LLMs & Yes & LLMs for materials science\\\midrule
     \citet{Peivaste2025} & Broad & Broad & No & Emphasis on data \\\midrule
     \citet{Jiang2025} & Text‑centric & Text‑centric & Yes & Literature mining \\\midrule
     \citet{lei2024materials} & Broad & Broad & No & LLMs augmenting human researchers \\\midrule\midrule

     \textbf{Ours} & Broad & Broad & Yes & End-to-end and pipeline-centric perspective \\\bottomrule
    \end{tabular}
    }
    \caption{Comparison of our survey with other most recent ones on AI (especially LLMs) for materials science.
    ``Broad'' in the Materials Domain column means the survey covers a wide range of materials science tasks, workflows, and concepts. 
    ``Broad'' in the AI Methods column means the survey covers most AI methods, including traditional machine learning, deep learning, and LLMs. 
    ``Discovery-Oriented'' is defined as systems prioritizing closed-loop validation and outcome-driven objectives.
    }
    \label{tab:comparison}
\end{table}

Through the \textit{pipeline-centric} lens, as shown in Fig. \ref{fig:overall}, this survey advocates shifting AI4MS from static fine-tuning to goal-oriented agent workflows for materials discovery. 
The discussion is structured around three guiding questions: \textbf{Q1} (Evolution of Agency): How have general-purpose machine learning models evolved from merely passive processors to agentic systems? \textbf{Q2} (Limitations of Phase-Centric Approaches) Why are existing AI for materials science systems not enough to meet the needs of autonomous materials discovery, even though they score well on material-related benchmarks? \textbf{Q3} (The End-to-End Gap) What capacities have been developed, and how can we bridge the gaps in capability, optimization, and governance between existing AI for materials science and an end-to-end autonomous materials discovery pipeline?

There exist surveys on relevant topics, but they differ from ours, which are summarized in Tab. \ref{tab:comparison}.
In contrast, we present a unified pipeline-centric perspective for review and discussion.
Driven by recent enablers such as autonomous labs, agent frameworks, and foundation models, this survey charts a practical roadmap from foundation models to autonomous, safety-aware LLM agents for discovering novel and useful materials that satisfy real performance, cost, and reliability criteria.
Targeting both AI researchers and materials scientists, we bridge these disciplines to show why a system-level paradigm shift is now both necessary and feasible.

\begin{figure}[t]
    \centering
    \includegraphics[width=0.75\linewidth]{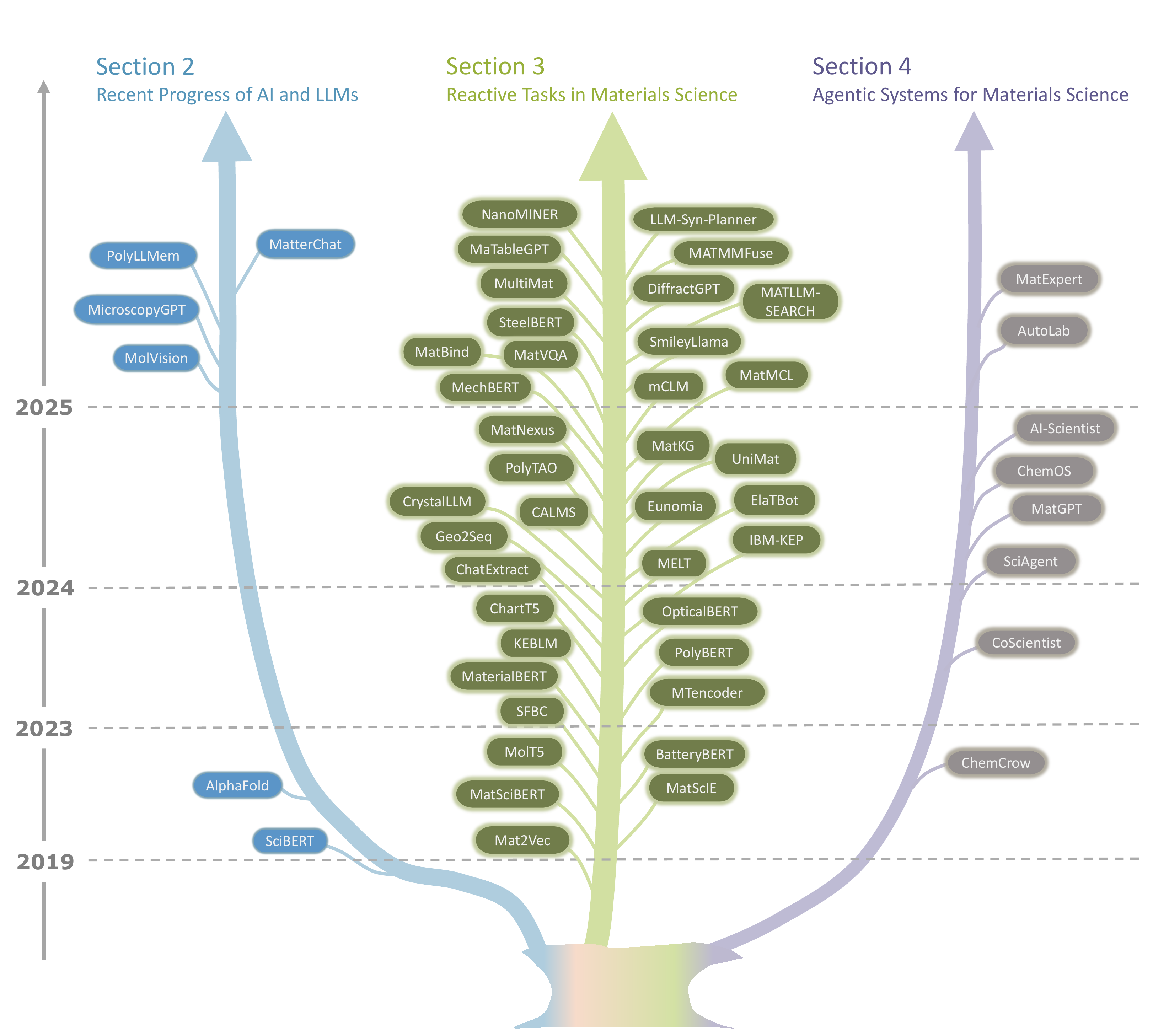}
    \caption{Technology tree of AI4Material science research. With the emergence of LLMs and agents, research on materials science initially focused on domain specific task, primarily concentrating on seperate reactive tasks. Subsequent research has delved deeper, gradually integrating more with agentic systems for materials science.}
    \label{fig:tree}
\end{figure}

\section{Recent Progress of AI and LLMs\label{sec:2_progress_of_AI}}

The evolution from traditional machine learning to LLMs has progressively recast AI from a computational tool into an intelligent research collaborator capable of reasoning, generating hypotheses, orchestrating simulations, and designing experiments~\citep{pyzerknapp2022accelerating}.
In this section, we review recent AI and LLM advances with prospective relevance to materials science—irrespective of current deployment—to address Q1 in Sec. \ref{sec:intro}: How have general-purpose machine learning models evolved from merely passive processors to agentic systems?
Rather than a chronological history, we trace this evolution through four key stages that mirror the pipeline in Fig. \ref{fig:overall}: (1) predictive models trained on static datasets; (2) foundation models serving as programmable priors; (3) post-training methods for objective shaping and controllability; and (4) agentic systems enabling tool use and long-horizon rewards.
As shown in Fig. \ref{fig:overall}, pre-trained foundation models generally initialize the agentic pipeline for materials science; consequently, pre-training choices shape downstream capabilities and should be iteratively adapted using RL signals from real‑world experiments.

\begin{figure}
    \footnotesize
    \begin{forest}
        for tree={
            forked edges,
            grow'=0,
            draw,
            rounded corners,
            node options={align=center,},
            text width=2.7cm,
            s sep=6pt,
            calign=child edge, calign child=(n_children()+1)/2,
            font=\small,
        },
        [Recent Progress of AI~\S\ref{sec:2_progress_of_AI}, fill=gray!45, parent
            [Deep Learning Revolution~\S\ref{sec:2_1}, for tree={ pretrain}
                [Machine Learning Foundations~\S\ref{sec:2_1_1}, for tree={ pretrain}
                    [
                        {\tstyle{Supervised learning}: e.g., \citet{Cortes1995}, \citet{Breiman2001}, \citet{Anderson1995}, }, pretrain_work
                    ]
                    [
                        {\tstyle{Unsupervised learning}: e.g., \citet{McQueen1967}, \citet{Khan2014}, \citet{Abdi2010}, \citet{Maaten2008}}, pretrain_work
                    ]
                    [
                        {\tstyle{Reinforcement learning}: e.g., \citet{Sutton1998}, \citet{Mnih2013}, \citet{Schulman2017}, \citet{Garcia1989}, \citet{Konda1999}}, pretrain_work
                    ]
                ]
                [Deep Learning~\S\ref{sec:2_1_2}, for tree={ pretrain}
                    [{\tstyle{Architecture}: e.g., \citet{Krizhevsky2012}, \citet{goodfellow2014generativeadversarialnetworks}, \citet{kingma2013auto}, \citet{computation2016long} }
                    , pretrain_work]
                    [{\tstyle{Training}: e.g., \citet{srivastava2014dropout}, \citet{ioffe2015batch}, \citet{abadi2016tensorflow}, \citet{paszke2019pytorch}}
                    , pretrain_work]
                ]
            ]
            [Transformer and Foundation Models~\S\ref{sec:2_2}, for tree={fill=red!45,template}
                [Attention Revolution~\S\ref{sec:2_2_1},  template
                    [{e.g., \citet{Vaswani2017}, \citet{Devlin2019}, \citet{brown2020language}}
                    , template_work]
                ]
                [Foundation Models~\S\ref{sec:2_2_2},  template
                    [{\tstyle{General-purpose}: e.g., \citet{Bommasani2021Foundations}};
                    , template_work]
                [{\tstyle{Science-specific}: e.g., \citet{beltagy-etal-2019-scibert}, \citet{Jumper2021highly}};
                    , template_work]
                ]
                [LLMs and Reasoning~\S\ref{sec:2_2_3},  template
                    [{\tstyle{General-purpose}: e.g., \citet{wei2022chainofthought}, \citet{borgeaud2022improving}, \citet{Kaplan2020ScalingLF}}
                    , template_work]
                    [{\tstyle{Materials-specific}: e.g., \citet{yu2024large}, \citet{wei2024crystal}, \citet{kim2021generative}  
                    }
                    , template_work]
                ]
                [Multimodal Integration~\S\ref{sec:2_2_4},  template
                    [{\tstyle{General-Purpose}: e.g., \citet{li2020multimodal}, \citet{radford2021learning}, \citet{li2022blip}}
                    , template_work]
                    [{\tstyle{Materials-specific}: e.g., \citet{Tang2025}, \citet{Choudhary2025}, \citet{Adak2025}}
                    , template_work]
                ]
                [Novel Architectures~\S\ref{sec:2_2_5},  template
                    [{e.g., \citet{gu2021efficiently}, \citet{gu2024mamba}}
                    , template_work]
                ]
            ]
            [LLMs and Agents~\S\ref{sec:2_3}, for tree={fill=blue!45, answer}
                [Emerging Paradigms~\S\ref{sec:2_3_1}, answer
                    [{e.g., \citet{ouyang2022training}, \citet{radford2021learning}
                    }
                    , answer_work]
                ]
                [Rise of Agentic System~\S\ref{sec:2_3_2}, answer
                    [{e.g., \citet{Yao2022}, \citet{Schick2023Toolformer}, \citet{Gravitas2023}
                    }
                    , answer_work]
                ]
                [Recent Advances in Agents~\S\ref{sec:2_3_3}, answer
                    [{e.g., \citet{anthropic2025claude37}, \citet{microsoft2025ai_agents}, \citet{jiang2025agenticscimlcollaborativemultiagentsystems}
                    }
                    , answer_work]
                ]
            ]
        ]
    \end{forest}
    \caption{Taxonomy of recent progress of AI and LLMs.}
    \label{tree:components}
    \end{figure}
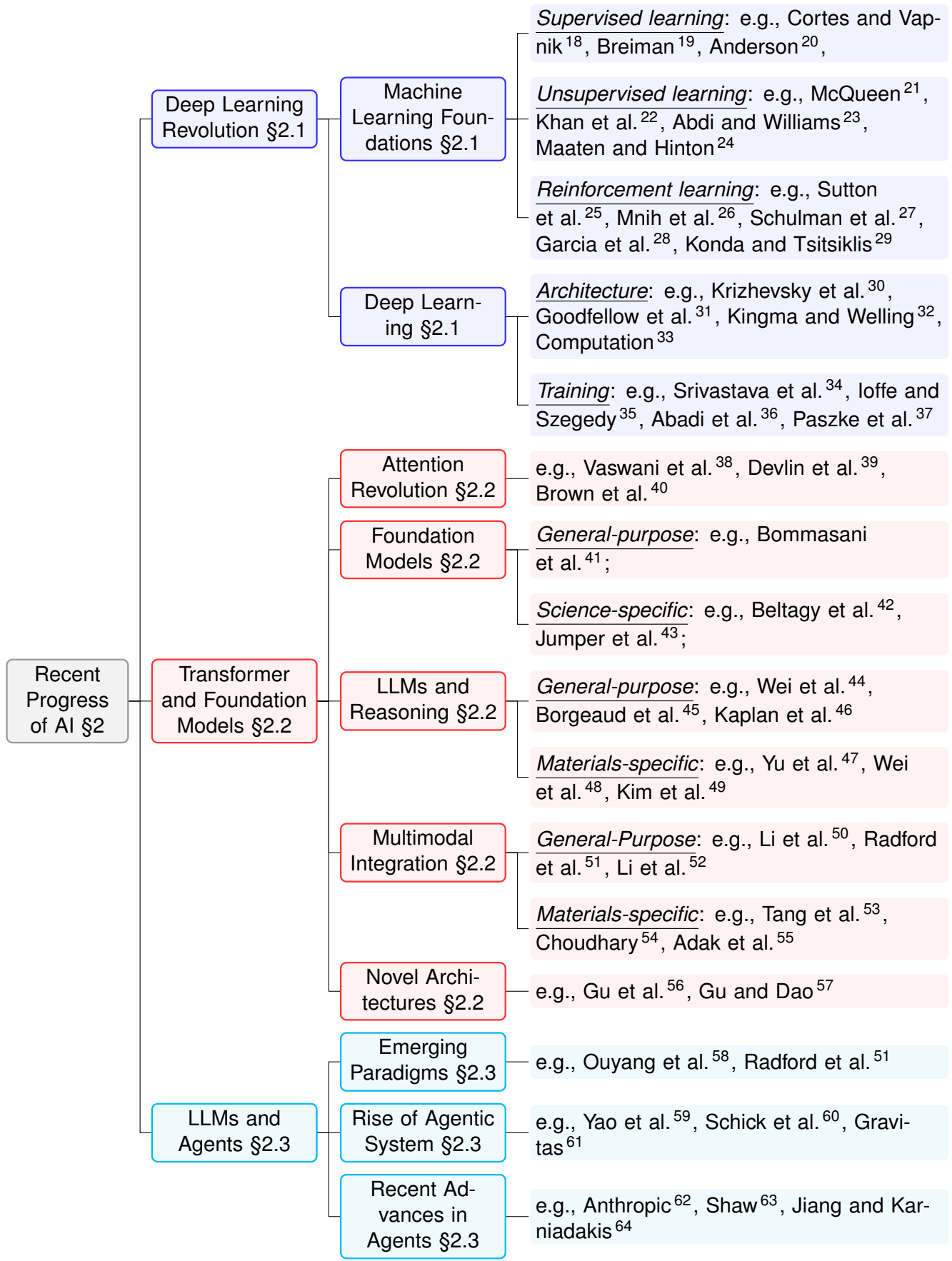

\subsection{Deep Learning Revolution}\label{sec:2_1}

\paragraph{Machine Learning Foundations}\label{sec:2_1_1}

From the late 1990s through the early 2010s, most early work in machine learning and AI was based on feature engineering, establishing the first batch of data-driven tools for materials science. 
Over this period, these early methods were not just used in materials research; they quickly became important instruments for finding new materials and predicting their properties.

There are three main types of machine learning, which have been applied to materials science. 
Supervised learning techniques, such as support vector machines~\citep{Cortes1995}, random forests~\citep{Breiman2001}, and shallow neural networks~\citep{Anderson1995}, facilitated property prediction from labeled data, thereby creating quantitative structure-property relationships for expedited materials screening. 
~\citet{Ward2016General} introduced a general-purpose framework that combined a chemically diverse feature set with ensemble-based models like random forests. 
Their framework worked well on both crystalline and amorphous systems, showing how powerful systematic feature design can be in speeding up the discovery of new materials.
Unsupervised learning techniques, such as clustering (e.g., K-Means~\citep{McQueen1967}, DBSCAN~\citep{Khan2014}) and dimensionality reduction (e.g., PCA~\citep{Abdi2010}, t-SNE~\citep{Maaten2008}), aided materials discovery by enabling pattern recognition, phase identification, and feature extraction from high-dimensional data~\citep{Ximendes2022Less}. 
Both supervised and unsupervised learning algorithms enabled the initial requirements of the discovery cycle, i.e., correlating structures to properties. 

To meet the need for autonomous materials discovery, RL techniques (e.g., Q-Learning~\citep{Sutton1998}, Deep Q-Networks (DQN)~\citep{Mnih2013}, Proximal Policy Optimization (PPO)~\citep{Schulman2017}, Model Predictive Control (MPC)~\citep{Garcia1989}, and Actor–Critic methods~\citep{Konda1999}) were introduced to interact with an environment and learn from the consequences of their actions by optimizing reward functions over sequences of decisions. These methods make it possible to explore chemical space~\citep{Kim2024Materials} and to search large design spaces in a more systematic way to identify materials with desirable properties~\citep{Choudhary2022Recent}. These models work together to form an important foundation for the deep learning architectures that emerged later.
While these classical and early RL methods provided the foundational logic for data-driven research, they lacked the representation learning capabilities required to process multi-modal experimental data.

\paragraph{Deep Learning}\label{sec:2_1_2}

The 2010s saw a shift away from feature engineering and toward representation learning. 
The availability of large-scale datasets and increased computational resources powered the development of deep learning~\citep{LeCun2015DeepLearning,Goodfellow-et-al-2016}. 
The 2012 ImageNet competition winner, AlexNet~\citep{Krizhevsky2012}, helped establish deep learning as a general-purpose machine learning tool for real-world problems. 
In materials workflows, this period matters less as a milestone in "AI history" and more as the point when learned representations started to replace manually designed descriptors~\cite{xie2018crystal,Ward2016General}.
This shift enabled scalable screening, surrogate modeling, and downstream decision-making in materials discovery pipelines~\cite{jha2018elemnet,butler2018machine}.

After discriminative models proved useful and matured, the focus shifted to generative modeling. Generative Adversarial Networks (GANs) \citep{goodfellow2014generativeadversarialnetworks} made inverse design possible by training generator and discriminator networks against each other. Variational Autoencoders (VAEs)~\citep{kingma2013auto} had efficient inference and learning in probabilistic generative models with continuous latent variables.

Recurrent architectures such as Long Short-Term Memory (LSTM)~\citep{computation2016long} and Gated Recurrent Units (GRU)~\citep{cho2014learning} have made temporal modeling for reaction prediction, synthesis pathway optimization, and materials degradation forecasting possible.
Gradient-based optimizers, such as Adam and SGD with momentum, together with regularization techniques like dropout~\citep{srivastava2014dropout} and batch normalization~\citep{ioffe2015batch} improved training stability and reliability.
Tools like TensorFlow~\citep{abadi2016tensorflow} and PyTorch~\citep{paszke2019pytorch} played an important role in simplifying model development. Especially their support for multi-GPU training further accelerated progress and made it feasible to work with large databases of materials~\citep{kailkhura2019reliable}.
As datasets continued to expand and recurrent architectures struggled with long-range dependencies, the community has explored models better suited to handle global context. 
This shift has led to the rise of the attention-based Transformer era.

\subsection{Transformer and Foundation Models}\label{sec:2_2}

\paragraph{The Attention Revolution}\label{sec:2_2_1}
The self-attention mechanism~\cite{Vaswani2017} has reshaped sequence modeling tasks since it was proposed in the Transformer framework. 
This made it possible to perform parallel computation and model long-range dependencies, which are important for understanding chemical SMILES strings~\cite{weininger1988smiles} and scientific literature. 
The multi-head attention~\cite{Vaswani2017} method enables models to look at many representation subspaces simultaneously, capturing complex interactions across them.

BERT (Bidirectional Encoder Representations from Transformers)~\citep{Devlin2019} introduced bidirectional pre-training with masked language modeling and next sentence prediction in late 2018.  
This enabled models to capture the complex, context-dependent relationships that characterize scientific writing.
GPT-3 \citep{brown2020language} demonstrated remarkable few-shot learning abilities with 175 billion parameters. Its performance suggested that scale, together with data, can give models surprising flexibility, even without heavy fine-tuning.

\paragraph{Foundation Models and Scale-Up}\label{sec:2_2_2}

The term ''foundation models'' was first used in \cite{Bommasani2021Foundations} to describe large models trained on a variety of data, typically using self-supervision at scale, that can be applied to a wide range of tasks. This is because attention-based architectures made it possible for these models to be built. 
These models represent a major shift in AI: their unprecedented size leads to new abilities in language, vision, reasoning, and human interaction, but they also increase the risks of homogenization, where flaws or biases in the foundation models spread to all adapted downstream systems.  
Task-specialized LLMs trained or adapted on curated corpora (e.g., code~\cite{chen2021evaluating,Nijkamp2022CodeGen}, math~\cite{lewkowycz2022solving,cobbe2021training}, legal text~\cite{chalkidis2020legal}) exemplify the dual nature of scaling: enhanced domain proficiency frequently accompanies hallucination~\cite{ji2023survey} and challenges in calibration and faithfulness~\cite{Li2022FaithfulnessIN}, alongside difficulties in controllability and verifiability~\cite{dziri2023faithfatelimitstransformers}.

Field-specific foundation models have also emerged for scientific use. 
Domain-adapted scientific foundation models tailor general priors to expert corpora and evaluation regimes.
SciBERT \cite{beltagy-etal-2019-scibert} modified BERT for scientific literature by training it on 1.14 million papers from Semantic Scholar. 
By improving skills like named entity recognition, citation prediction, and relation extraction, it made biomedical and computer science texts easier to understand.
AlphaFold \cite{Jumper2021highly} marked a major advance in predicting protein structures with attention mechanisms for evolutionary data and physical constraints. 
This showed how transformers can be used for complex scientific problems, motivating analogous designs in other scientific domains (including materials science).

\paragraph{LLMs and Reasoning Capabilities}\label{sec:2_2_3}
Recent research has focused on reasoning, knowledge grounding, and tool-enhanced inference in addition to scaling.
Chain-of-thought prompting \citep{wei2022chainofthought} adds the scientific reasoning that is essential for materials discovery.
Memory-augmented architectures \citep{borgeaud2022improving} support the retrieval of external knowledge, hence enhancing access to specialist materials science literature and databases during inference. 
Scaling laws \citep{Kaplan2020ScalingLF} show that performance improves when the number of parameters and data increases, following power-law relationships up to the limits of the regime studied. 
This poses a challenge for materials research because high-quality datasets are far less abundant than in many other domains.

In response to data scarcity, the community has explored strategies including the use of synthetic data such as OC20~\cite{Chanussot_2021_OC20}, pre-training on large unlabeled materials databases such as the Materials Project~\cite{jain2013materials}, and transfer learning from related scientific domains, including chemical datasets such as QM9~\cite{ramakrishnan2014quantum}.
The inherent tension between data requirements and availability constrains applications of foundation models in materials discovery to subdomains with sufficient training instances. 
The improvement in the reasoning ability of LLMs enhanced multiple fields in materials science research~\citep{yu2024large}, such as enhanced reasoning, which can help molecular structure design~\citep{kim2021generative} and materials design~\citep{wei2024crystal}.

\paragraph{Multimodal Integration}\label{sec:2_2_4}

Cross-modal attention mechanisms have been used in multimodal foundation models \citep{li2020multimodal,singh2022flava} to integrate information from text, images, and structured data, which is also useful for materials science, when workflows require fusing heterogeneous evidence (e.g., literature, microscopy, and structure representations) into a single decision context. 
Vision-language models (VLMs) like CLIP \cite{radford2021learning} and BLIP \cite{li2022blip} use contrastive learning to enhance vision and language representation alignment and thereby provide retrieval-ready joint embeddings that can be repurposed with limited supervision. 
This ability has enabled tasks such as zero-shot classification and text-image retrieval for the materials characterization study (e.g., rapid triage, annotation, and cross-referencing of characterization images with textual descriptors), but reliability typically depends on domain-specific calibration and evaluation.    
Flamingo \cite{alayrac2022flamingo}, KOSMOS-2 \cite{peng2023kosmos2}, and VILA \cite{lin2024vila} are working on enhancing the video and 3D understanding abilities of the current multimodal framework, by extending multimodal context length and enabling richer image–text interleaving; however, robust 3D understanding for scientific data streams remains an open gap for end-to-end discovery settings.

In addition to general-purpose multimodal models, there are also specialized multimodal LLMs that have been adapted for materials science. These include MatterChat~\citep{Tang2025}, MicroscopyGPT~\citep{Choudhary2025}, MolVision~\citep{Adak2025}, and PolyLLMem~\citep{Zhang2025}. These models combine atomic structure data, text, and images to make it easier for people and AI to work jointly on tasks such as discovering new materials and predicting their properties. 
However, the impact on discovery depends on how well the system integrates aspects like tool use, uncertainty estimation, and closed-loop validation, not just the model's ability in isolation.

\paragraph{Novel Architectures Beyond Transformers\label{sec:novel_architecture}\label{sec:2_2_5}}

To address the computational efficiency bottleneck of Transformer-based architectures on long scientific sequences, several alternative architectures have been proposed.
Structured State Space Models (S4) \citep{gu2021efficiently} addressed the limitations of time-series materials data and molecular dynamics simulations by implementing parameter-efficient architectures that facilitated training on exceptionally lengthy sequences (16K+ tokens).  
Mamba \citep{gu2024mamba}, a selective state space model with input-dependent state transitions and linear computing complexity, can reach Transformer-level quality on language-modeling benchmarks while scaling more favorably with sequence length.
This efficiency is particularly relevant when models must handle long molecular strings (e.g., SMILES or SELFIES for complex molecules and polymers), simulation trajectories from molecular dynamics or density functional theory (DFT) relaxations, lab notebooks accumulated over long experimental campaigns, or tool outputs as part of an end-to-end discovery workflow.
The next paradigm shift moves away from passively reading text and toward reasoning and tool usage skills, leading to autonomous agents.

\subsection{LLMs and Agents\label{sec:2_llm_agent}\label{sec:2_3}}

The recent innovations in AI have introduced what is termed ``The Era of Experience''~\citep{silver2025era}.  
In this view, progress is driven not only by fitting static human-generated corpora, but also by iterative interaction with environments that provide feedback signals.
Now the systems-based LLM can complete complex experimental tasks to be an agent rather than a single network predictor. By interacting with the environment/workflow, the agent can learn, improve, and adapt to real-world tasks.

The AI field is currently transitioning from the ``Era of Human Data,'' where progress relies on imitating finite, static datasets, to the ``Era of Experience,'' where agents achieve superhuman capabilities by actively interacting with their environments and learning from the consequences of their actions~\citep{silver2025era}. 
The relevance of this shift to materials science is practical: the interactive nature can steer the whole pipeline towards the ultimate goal of discovering novel and useful materials.

Just as general-purpose agents are now moving beyond human-privileged text interactions to autonomous streams of experience, AI for materials science must evolve from passive models trained on historical property databases to active AI Scientist~\citep{Wang2023NatureAIDiscovery} agents that engage in closed-loop discovery. 
By treating experiments and simulations not merely as data sources but as interactive environments, such agents can generate their own high-quality data through RL, potentially expanding exploration beyond human-selected candidates and accelerating iterative design–test–learn cycles.

\paragraph{Emerging Paradigms}\label{sec:2_3_1}

New paradigms, such as alignment and instruction tuning (e.g., Constitutional AI~\citep{bai2022constitutional}, RLHF~\citep{ouyang2022training}), have improved instruction following and controllability and can reduce hallucinations in practice, although factuality is not guaranteed and typically depends on grounding and verification, especially for scientific domains like materials science. 
These paradigms are based on foundation models and can be viewed as post-training objective shaping on top of pre-training.
Few-shot~\citep{finn2017model} and zero-shot learning~\citep{radford2021learning} have made it easier to adapt to tasks with little data. This is important in fields where labeled datasets are limited by the high cost of experiments and the need for domain expertise. 
In-context learning~\citep{brown2020language, garg2022incontext} enables rapid prototyping of new actions directly from stimuli, removing the need for task-specific fine-tuning.
In order to improve scientific decision-making rather than just isolated perception, models are becoming more skilled at tool- and evidence-grounded reasoning across a variety of modalities, such as images~\citep{chen2024interpretable}, crystal structures~\citep{Xie2018}, spectroscopic data, and experimental measurements. 
These works suggest that agentic systems couple language-based reasoning with external actions (e.g., retrieval, simulation, and code execution).

\paragraph{Rise of Agentic Systems}\label{sec:2_3_2}
Enhanced agentic systems have been developed during the evolution of LLMs.
In order to make decisions and improve grounding, agent loops combine reasoning and action by utilizing external tools such as retrieval~\citep{lewis2020retrieval}, calculators and symbolic computation~\citep{gao2023pal}, code execution or function calling~\citep{Yao2022}, browsers~\citep{Nakano2021WebGPTBQ}, structured databases or enterprise APIs~\citep{qin2023toolllm}. 
Later frameworks like Toolformer~\citep{Schick2023Toolformer}, AutoGPT~\citep{Gravitas2023}, and LangChain~\citep{Chase2022} used these features to add persistent memory, the ability to work with more than one tool at a time, and the ability to plan recursively. 
These works demonstrate that agentic systems can address problems necessitating extended procedures, although consistent long-term performance continues to be influenced by tool reliability, feedback design, and safety constraints.

\paragraph{Recent Advances in Agents}\label{sec:2_3_3}
 
In the second half of 2025, a wave of agents with thinking and embodied action have emerged.
Anthropic's Claude Sonnet~\citep{anthropic2025claude37} is a cognitive agent equipped with enhanced reasoning, understanding, and tool-using abilities.
This is a step toward AI systems that are better at using tools and sustaining multi-step analysis in real workflows.
It also supports more scientific use by helping users structure questions, integrate evidence, and connect ideas across domains.
Gemini Robotics 1.5~\citep{Gemini_Robotics_Team2025-qt} demonstrated transfer learning across different robotic embodiments, advancing robotic agents beyond single-platform policies.
Vision-language-action (VLA) models highlight the value of explicit intermediate plans or reasoning steps that can improve action.
NVIDIA's Cosmos reasoning model~\citep{nvidia2025cosmosreason1physicalcommonsense} improves agent reasoning by integrating physical priors and common sense constraints into robotic decision-making.
Microsoft's autonomous agents~\citep{microsoft2025ai_agents} demonstrate that AI agents have great potential to orchestrate complex, multi-step tasks on behalf of users.

Through the above improvements, multiple agents with different skills can collaborate to build agent ecosystems capable of completing complex tasks.
Recent systems demonstrate that AI can function as a powerful research partner, able to read literature, generate hypotheses, plan experiments, and design syntheses, rather than just making predictions~\citep{Bran2024NatureChemCrow}. 
Such agents can retrieve external resources, call specialized tools, and coordinate downstream actions autonomously~\citep{lu2024aiscientistfullyautomated,Gravitas2023}.   

Integrating data-driven analysis, such as extracting useful information from large-scale scientific literature, with formulating hypotheses and organizing experiments can be supported by LLM-based agents.
Such capabilities are discussed in detail in~\citep{jiang2025agenticscimlcollaborativemultiagentsystems}.
These features make it much easier to sort data for tasks that only apply to a certain area. 

However, end-to-end materials discovery is still limited by reliable grounding, constraint-aware reasoning such as stability and synthesizability, and closed-loop validation with simulations and experiments.
Agents can also enable more independent, agent-driven research pipelines that may accelerate discoveries by emphasizing system-level integration (feedback, coordination, and objective alignment) rather than isolated model capabilities.

\subsection{Pipeline-Centric Perspective\label{sec2:pipeline-perspective}}

\begin{figure}[ht]
    \centering
    \includegraphics[width=0.9\textwidth]{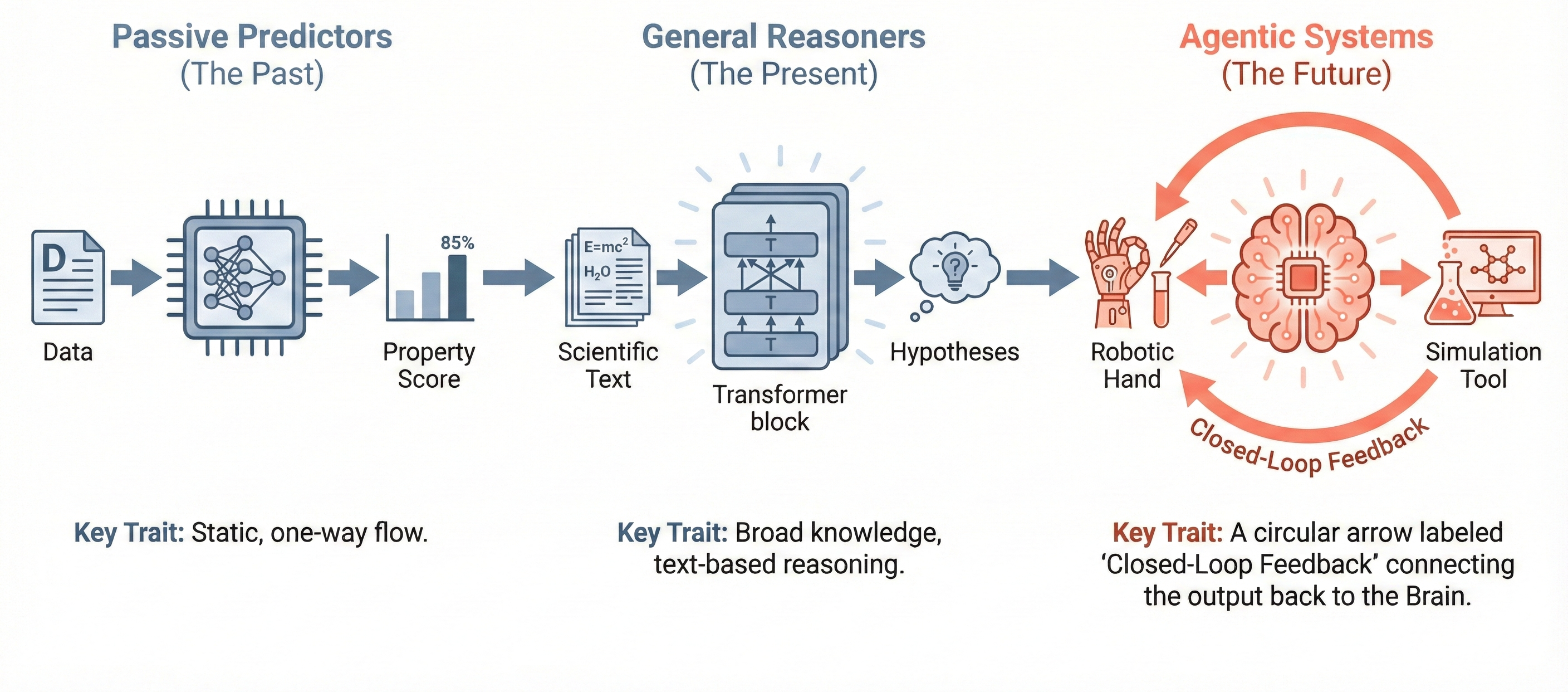}
    \caption{The gradual integration of general AI into the materials science workflow. This framework illustrates the progressive evolution of broad, pre-trained foundation models as they are adapted for scientific discovery. Rather than a sudden replacement, the diagram depicts how general capabilities are iteratively refined through domain adaptation and feedback loops, allowing general-purpose AI to be gradually and effectively applied to specialized materials challenges.
    }
    \label{fig:sec2_perspective}
\end{figure}

After addressing Q1, we discuss the interaction between the pre‑trained LLMs and the agentic LLMs, as shown in Fig. \ref{fig:sec2_perspective}, from the end‑to‑end pipeline‑centric perspective, explicitly treating the pre-training stage as a controllable module whose data, objectives, and safety constraints can be continuously revised in light of downstream discovery performance rather than as a one‑off static initialization. 
This framing makes pre-training an object of system design—subject to iterative revision—rather than a fixed prerequisite that is optimized once and then taken for granted.

It is generally believed that general pre-training corpora on which contemporary LLMs are trained are essential to the downstream fine-tuned models; however, recent studies (e.g., \citep{JolicoeurMartineau2025}) suggest that such broad pre-training may not always be necessary and, in some cases, may even be suboptimal for specialized scientific applications. Furthermore, reliance on indiscriminately scraped web‑scale corpora increases the chance of poisoning attacks and incidental misinformation, which can induce harmful or misleading behaviors even when benchmarks appear unaffected \citep{Souly2025}. Instead, if the ultimate objective of discovering novel materials is what we value most, we should review the techniques presented before and enforce that every upstream design choice—including corpus selection, pre-training objectives, and alignment procedures—is justified by its contribution to this end goal rather than by improvements on proxy language benchmarks. 
More concretely, better pre-training should be defined by downstream discovery-relevant behaviors (e.g., planning feasible syntheses and correctly handling hazards), not by generic language metrics alone.
Continuing the presentations in Sec. \ref{sec:novel_architecture}, we should require that these architectures benefit the discovery of novel materials and not just improve the pre‑training scores on generic benchmarks, recognizing that even small fractions of poisoned or misaligned data can strongly bias downstream scientific behavior.

To make this pipeline‑centric stance operational, the pre-training stage must be embedded within a feedback loop in which agentic systems operating over simulations and experiments supply long‑horizon reward signals that can, in principle, be propagated back to earlier stages to revise corpus composition, sampling strategies, and alignment objectives whenever inherited behaviors of a general‑purpose LLM systematically hinder materials discovery. This suggests moving from a frozen pre-training corpus toward an adaptive, feedback‑driven pre-training loop where materials‑science outcomes and safety considerations jointly determine which segments of the original web‑scale data are amplified, attenuated, or removed, and where domain‑specific corpora are dynamically expanded in response to identified capability gaps along the design–synthesis–validation pipeline. Recent progress in LLM‑based agent architectures and tooling, as discussed in Sec. \ref{sec:2_llm_agent}, makes such a pipeline‑centric perspective increasingly feasible, allowing the successful discovery of novel, experimentally validated materials to serve as the primary reward signal guiding the entire system rather than surrogate benchmarks alone.
Crucially, this closes the evaluation loop: it ties what we optimize (training choices) to what we ultimately value (validated discovery).

Influence functions \citep{Li2024-yp,Grosse2023-cu} offer one concrete mechanism for such backward credit assignment by estimating how infinitesimal upweighting or downweighting of individual training examples would perturb a model’s parameters and downstream predictions. 
Approximate influence estimators \citep{Li2024-yp} based on scalable inverse‑Hessian approximations have been demonstrated on large LLMs, recovering sparse, interpretable sets of pre-training sequences that are most responsible for particular emergent behaviors. 
From a pipeline‑centric perspective, these tools can be repurposed to identify which general‑domain documents and which MatSci‑specific corpora most strongly support or undermine key capabilities needed for discovery (e.g., robust reasoning over phase diagrams, synthesis planning under realistic process constraints, or safety‑critical hazard identification), thereby informing which data sources to emphasize, de‑emphasize, or excise in subsequent rounds of pre-training.
This turns data selection into a goal-driven hyperparameter, rather than a one-time curation decision.

Analogously, an AI4MatSci pipeline could employ influence scores computed on discovery-relevant behaviors, e.g., successful identification of synthesizable candidates, safe handling of hazardous chemistries, or robust performance in long‑horizon design loops. 
It can upweight beneficial segments of both general and MatSci‑specific data, while downweighting or removing sequences that systematically induce hallucinated mechanisms or impossible synthesis conditions. 
In this view, pre-training, domain adaptation, and instruction tuning collectively form a single, influence‑addressable memory that is continuously reshaped by reward signals originating from agentic closed loops in simulation and experiments, rather than being optimized only once against proxy language benchmarks that may fail to reveal poisoning or subtle misalignment. 
In this sense, influence functions become not only interpretability tools but also prospective levers for dynamically re‑aligning the pre-training component with the ultimate materials‑science objective of discovering novel, useful, and safe materials, while mitigating security risks introduced by open‑world data collection. 
At the same time, these general‑purpose agents must be carefully tailored to materials science to align with real workflows, safety regulations, and experimental constraints, a refinement that will be developed in the next section.

\section{Reactive Tasks in Materials Science from an AI Perspective\label{sec:3_AI_for_matsci}}

\begin{figure}
    \footnotesize
    \begin{forest}
        for tree={
            forked edges,
            grow'=0,
            draw,
            rounded corners,
            node options={align=center,},
            text width=2.7cm,
            s sep=6pt,
            calign=child edge, calign child=(n_children()+1)/2,
            font=\small,
        },
        [Reactive Tasks in AI for Materials Science~\S\ref{sec:3_AI_for_matsci}, fill=gray!45, parent
            [Prediction\\~\S~\ref{sec:3_prediction}, for tree={child}
                [Regression\\~\S~\ref{sec:3_1_regression}, for tree={pretrain}
                    [
                        {\tstyle{Electronic Properties}:e.g., \citet{jin2024crystal}, \citet{jin2025transformer}, \citet{li2023materials}}, pretrain_work
                    ]
                    [
                        {\tstyle{Mechanical Properties}: e.g., \citet{lee2022evaluation}, \citet{liu2024large}}, pretrain_work
                    ]
                    [
                        {\tstyle{Thermodynamic Stability}: e.g., \citet{faber2015crystal}, \citet{houchins2024formation}}, pretrain_work
                    ]
                    [
                        {\tstyle{Thermoelectric and Thermal Properties}: e.g., \citet{ghosh2025automated}, \citet{korop2025application}}, pretrain_work
                    ]
                ]
                [Classification\\~\S~\ref{sec:3_1_classification}, for tree={ pretrain}
                    [{\tstyle{Defect Classification and Experimental Automation}: e.g., \citet{bran2024chemcrow}, \citet{boiko2023emergent}}
                    , pretrain_work]
                ]
                [Advanced Methodologies\\~\S\ref{sec:3_1_3_adv}, for tree={ pretrain}
                    [{\tstyle{Hybrid GNN-Transformer Architectures}: e.g., \citet{du2024ctgnn}, \citet{li2025hybrid}}
                    , pretrain_work]
                    [{\tstyle{Multi-Task Learning (MTL) for Generalization}: e.g., \citet{caruana1997multitask}, \citet{prein2023mtencoder}}
                    , pretrain_work]
                    [{\tstyle{Uncertainty Quantification}: e.g., \citet{tavazza2021uncertainty}, \citet{tavazza2021uncertainty}}
                    , pretrain_work]
                ]
            ]
            [Mining\\~\S\ref{sec:3_2_mining}, for tree={fill=red!45,template}
                [Information Extraction\\~\S\ref{sec:3_2_1_ie},  template
                    [{e.g., \citet{kim2017sdata_oxide}, \citet{cde_battery_2020_sdata}, \citet{sierepeklis2022thermo_sdata}, \citet{cde_stressstrain_2024_sdata}}
                    , template_work]
                ]
                [Knowledge Graph\\~\S\ref{sec:3_2_2_kg},  template
                    [{e.g., \citet{npj_alloy_semisup_2023}, \citet{npj_phase_diagrams_2020}, \citet{sdata_matkg_2024,sdata_termkg_2024}}
                    , template_work]
                ]
                [Database Automation and Annotation\\~\S\ref{sec:3_2_3_db},  template
                    [{e.g., \citet{sdata_auto_corpus_2022}, \citet{softwarex_matnexus_2024}, \citet{cms_matscie_2021}}
                    , template_work]
                ]
            ]
            [Generation\\~\S\ref{sec:3_3_generation}, for tree={fill=blue!45, answer}
                [Structure Generation\\~\S\ref{sec:3_3_1_stru}, answer
                    [{e.g., \citet{antunes2024crystalstructuregenerationautoregressive}, \citet{gruver2024finetuned}, \citet{gan2025large}
                    }
                    , answer_work]
                ]
                [Inverse Design\\~\S\ref{sec:3_3_2_id}, answer
                    [{e.g., \citet{takahara2025acceleratedinorganicmaterialsdesign}, \citet{pan2022deepreinforcementlearninginverse}, \citet{cleeton2025inverse}, \citet{fung2021inversedesigntwodimensionalmaterials}
                    }
                    , answer_work]
                ]
                [Synthesis Route Generation\\~\S\ref{sec:3_3_3_srg}, answer
                    [{e.g., \citet{mcdermott2021graph}, \citet{huo2022machine}, \citet{he2023precursor}
                    }
                    , answer_work]
                ]
            ]
        ]
    \end{forest}
    \caption{Taxonomy of reactive tasks in materials science from an AI perspective, Part 1.}
    \label{tree:secion-3-part1}
    \end{figure}
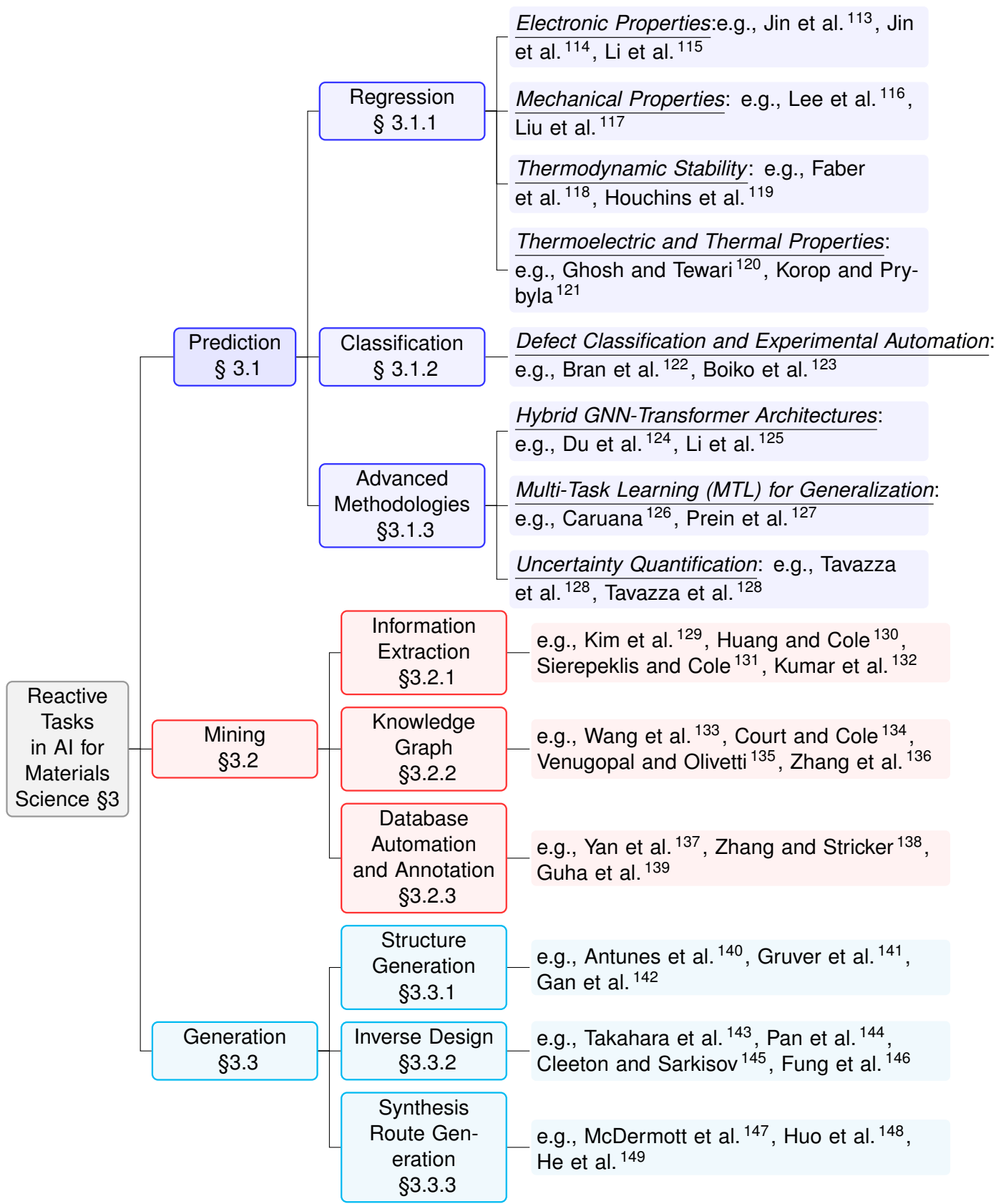

    \begin{figure}
        \footnotesize
        \begin{forest}
            for tree={
                forked edges,
                grow'=0,
                draw,
                rounded corners,
                node options={align=center,},
                text width=2.7cm,
                s sep=6pt,
                calign=child edge, calign child=(n_children()+1)/2,
                font=\small,
            },
            [Reactive Tasks in AI for Materials Science~\S\ref{sec:3_AI_for_matsci}, fill=gray!45, parent
            [Optimization and Verification\\~\S\ref{sec:3_4_optimization}, for tree={fill=blue!45, template}
                [Materials Discovery Process\\~\S\ref{sec:3_4_1_mdpo}, template
                    [{e.g., \citet{macleod2022self}, \citet{chitturi2024targeted}, \citet{zhang2023optimizing}
                    }
                    , template_work]
                ]
                [Simulation and AI-based Verification\\~\S\ref{sec:3_4_2_simulation}, template
                    [{e.g., \citet{fare2022multi}, \citet{merchant2023scaling}, \citet{lan2024enabling}
                    }
                    , template_work]
                ]
                [Agent-based Closed-loop Labs\\~\S\ref{sec:3_4_3_closed_lab}, template
                    [{e.g., \citet{kulichenko2024data}, \citet{abolhasani2023rise}, \citet{seifrid2022autonomous}
                    }
                    , template_work]
                ]
            ]
                [Data and Knowledge\\~\S\ref{sec:3_5_reactive_challenges_data}, for tree={fill=blue!45, answer}
                    [Data scarcity and Heterogeneity\\~\S\ref{sec:3_5_1_data_scarcity}, answer
                        [{\tstyle{Data Augmentation}: e.g., \citet{chanussot2021open}, \citet{szymanski2024integrated}, \citet{gibson2022data}}
                        , answer_work]
                        [{\tstyle{Multi-fidelity learning}: e.g., \citet{pilania2017multi}, \citet{debreuck2022accurate}}
                        , answer_work]
                        [{\tstyle{Few-shot Learning}: e.g., \citet{guo2021few}, \citet{qian2024meta}, \citet{zhang2024molfescue}}
                        , answer_work]
                    ]
                    [Knowledge integration\\~\S\ref{sec:3_5_2_knowledge_integration}, answer
                        [{\tstyle{Physics-informed Neural Networks}: e.g., \citet{chen2020physics}, \citet{fang2019deep}, \citet{zhang2022analyses}
                        }
                        , answer_work]
                    [{\tstyle{Structure-based Knowledge Integration}: e.g., \citet{jha2018elemnet}, \citet{wang2021compositionally}, \citet{chen2019graph}
                        }
                        , answer_work]
                    [{\tstyle{Ontology and Knowledge Graph Integration}: e.g., \citet{zhao2021knowledge}, \citet{fang2022molecular}, \citet{fang2023knowledge}
                        }
                        , answer_work]
                        [{\tstyle{Pre-trained Model}: e.g., \citet{huang2022batterybert}, \citet{zhao2023opticalbert}, \citet{ock2024unimat}
                        }
                        , answer_work]
                    ]
                    [Multimodality\\~\S\ref{sec:3_5_3_multimodality_challenges}, answer
                        [{\tstyle{Multimodal Scientific Data}: e.g., \citet{gupta2022matscibert}, \citet{m3gnet2022}, \citet{xrd2025}
                        }
                        , answer_work]
                        [{\tstyle{Multimodal Fusion}: e.g., \citet{moro2025multimodal}, \citet{pai2025reliability}
                        }
                        , answer_work]
                    ]
                ]
                [Explainability\\~\S\ref{sec:3_6_reactive_explain}, for tree={ pretrain}
                [Sparse and Closed-form Models~\S\\\ref{sec:3_6_1_sparse_model}, for tree={ pretrain}
                    [
                        {e.g., \citet{Ouyang2018_SISSO}, \citet{Ouyang2022_iSISSO}, \citet{TorchSISSO2024}}, pretrain_work
                    ]
                ]
                [Attention and Graph Explainers\\~\S\ref{sec:3_6_2_attention_graph}, for tree={ pretrain}
                    [{e.g., \citet{Jain2019_AttentionNotExplanation}, \citet{Kotobi2024_XAS}, \citet{Das2022_CrysXPP}}
                    , pretrain_work]
                ]
                [Physics-informed Interpretability\\~\S\ref{sec:3_6_3_physics}, for tree={ pretrain}
                    [{e.g., \citet{Batzner2022_E3NN}, \citet{martonova2025generalized}} %, \citet{Xie2024_Spectra}}
                    , pretrain_work]
                ]
            ]
            ]
        \end{forest}
        \caption{Taxonomy of reactive tasks in materials science from an AI perspective, Part 2.}
        \label{tree:secion-3-part2}
        \end{figure}

% ICME \citep{allison2006integrated} and MGI \citep{de2019new}
\citep{ramprasad2017machine} have emphasized the importance of leveraging AI to accelerate the discovery of advanced materials technologies by integrating information and decision-making across the composition-process-structure-property-performance chain.
Having discussed recent AI progress in the previous section, we will now review the evolution of traditional AI for materials science here to ultimately answer the second guiding question \textbf{Q2} introduced in Sec.~\ref{sec:intro}: \emph{Why are existing AI for materials science systems not enough to meet the needs of autonomous materials discovery, even though they score well on material-related benchmarks?}

We will describe the progress and contributions of AI methods across predominantly reactive materials science tasks.
However, we must carefully interrogate this progress in light of a pipeline-centric perspective. 
Benchmark performance on isolated tasks such as property prediction or information extraction can be misleading proxies for true discovery outcomes when viewed outside the context of an end-to-end discovery loop, where synthesis feasibility, safety constraints, and experimental validation ultimately determine success. 
As shown in Fig.~\ref{fig:overall}, contemporary materials-science-tuned LLMs serve as key building blocks of agentic systems and can be further specialized to improve agentic performance; yet, optimizing individual modules for high benchmark scores alone may not translate into more efficient autonomous discovery without pipeline-level objective alignment and credit assignment from real discovery outcomes back to upstream components.

\subsection{Prediction\label{sec:3_prediction}}

Applying LLMs to materials science marks a significant paradigm shift, moving predictive modeling from structure-specific deep learning toward generalizable, knowledge-intensive representation learning~\citep{lei2024materials}. 
However, most current deployments still operate in a largely reactive, task-conditioned manner.
Prediction in this context encompasses both quantitative forecasting of continuous properties (regression) and categorical assignment (classification), applied across diverse materials data modalities. We argue that the key innovation of applying LLMs for materials science lies in its dual capacity to process symbolic structural inputs (e.g., formulas~\cite{zhao2025developing}, space-group/Wyckoff descriptors~\cite{kazeev2025wyckoff}, and other text-encoded crystallographic representations~\cite{antunes2023crystal}) and high-dimensional feature sets (e.g., Crystal–structure–derived geometric features~\cite{song2025accurate}, DFT-derived features~\cite{fung2021benchmarking}, foundation-model embeddings~\cite{takeda2023multi}). 
This capability not only helps the development of predictive models with enhanced performance and multimodal adaptability but also facilitates the assessment of their trustworthiness~\cite{moro2025multimodal}.

\subsubsection{Regression Tasks\label{sec:3_1_regression}}

Regression tasks for materials predict continuous-valued physical properties, such as thermodynamic stability~\cite{schmidt2017predicting}, mechanical stiffness~\cite{hooshmand2023machine}, and electrical and thermal characteristics~\cite{zhang2017investigation}. This is a crucial step in accelerating the search for and design of novel materials. 
In practice, these tasks typically assume fixed property definitions and evaluation protocols, which can limit adaptability in end-to-end discovery pipelines~\cite{riebesell2025framework}.
The effectiveness of these prediction models relies on their contextually enriched embeddings, which are obtained from various representations of the materials, such as crystal structure~\citep{Xie2018}, chemical composition~\citep{goodall2020predicting}, or textual descriptions~\cite{tshitoyan2019unsupervised}. 
Standard machine learning and GNN models have established strong benchmarks. 
However, the deployment of LLMs has introduced improved capabilities for property prediction by utilizing global context and multi-modal data~\citep{lei2024materials}. 
Still, it has trouble adapting~\citep{alampara2024mattext} across domain shift, new target properties, and constraints relevant to synthesis and processing.

The performance of machine learning models critically depends on their input representations. 
This challenge was acutely evident in early models for periodic solids, which had difficulty formulating representations that were invariant to arbitrary choices of coordinate systems or unit cells while guaranteeing uniqueness~\cite{schutt2014represent}. 
Early approaches adapted representations from molecular machine learning (e.g., the Coulomb matrix) to periodic systems by incorporating Ewald sums or considering neighboring unit cells~\citep{faber2015crystal}.
A paradigm shift was subsequently established by GNNs through representing crystals as graphs with atoms and bonds as nodes and edges, which allows models to learn from the local atomic environment through message-passing~\cite{xie2018crystal}.
More recently, Transformer-based architectures have built upon this graph-based foundation by processing materials as sequences of tokens—whether atoms, sites, or textual descriptions—a strategy that, by focusing on global context, more effectively captures long-range interactions and symmetries than the locally oriented GNNs leading to improved modeling performance \citep{srinivas2024cross}.
However, improved representations alone do not resolve long-horizon planning, feedback coupling, or decision-level constraints that arise when predictors are embedded in closed-loop discovery workflows.

\paragraph{Electronic Properties }
In current benchmark-driven screening workflows, much emphasis has been placed on the prediction of fundamental electronic properties, especially the bandgap and conductivity for the screening of semiconductors and functional materials~\cite{dunn2020benchmarking}.
This focus is partly pragmatic: these targets are widely used as first-pass filters, but they can also over-represent what is easiest to score on proxy benchmarks.
Recent works based on Transformers have shown a superior capability in capturing global crystal characteristics compared to previous successes of GNNs at capturing localized atomic environments~\citep{jin2024crystal}. 
The CrystalTransformer model embodies this advance by feeding in crystal structure information through a Transformer encoder; specifically, this work starts with a concatenation of chemical and structural feature matrices, and then processes the combined representation into a Transformer encoder. 
In the encoder, the multi-head self-attention identifies complex, long-range correlations between all atoms in a manner that is agnostic to their spatial proximity, enabling the model to learn highly context-aware atomic embeddings~\citep{jin2025transformer}. 
Other variants, such as MatInFormer, have generalized this paradigm by learning the "grammar of crystallography" through the tokenization of space group information and subsequently demonstrating the adaptability of this architecture for property prediction on a range of datasets~\citep{li2023materials}. Increasingly, a key factor is the strategic use of linguistic input. 
For example, if natural language descriptions of a crystalline solid are used, then high-accuracy predictions of properties such as bandgap can be achieved, which seems not possible for some GNNs due to an inability to capture certain nuanced, global structural details such as space group symmetry and Wyckoff positions~\citep{rubungo2023llm}. 
The success of Transformer-based architectures in the processing of both symbolic structural information and text suggests that they operate as superior contextual encoders rather than purely topological encoders~\cite{tang2025matterchat}.
This allows them to implicitly integrate learned chemical and structural knowledge, avoiding the limitations of localized message-passing schemes.

\paragraph{Mechanical Properties }
The prediction of mechanical properties such as bulk modulus and shear modulus is important in materials design with targeted strength, durability, and resilience~\cite{jo2021machine, kandavalli2023design}. Various machine learning models have been widely adopted for predicting such mechanical properties, greatly reducing their dependence on simulation methods that are computationally intensive~\citep{lee2022evaluation}. 
More recently, domain-specific LLMs for materials science can be specialized in the predictions of complex mechanical properties, often by leveraging text-derived representations or hybrid text+structured inputs.
For example, ElaTBot is an LLM for the prediction of the full elastic constant tensor from text inputs~\citep{liu2024large}. Hence, the elastic tensor can be predicted simultaneously with the determination of properties such as bulk modulus at any finite temperature. 
Moreover, by incorporating general-purpose LLMs like GPT-4o and using the Retrieval-Augmented Generation technique, its prediction accuracy can be boosted, indicating again the power of combining domain-specific knowledge with the broad reasoning capabilities of generalist models when retrieval sources are curated and verifiable~\citep{llamp2024retrieval}.
This significantly accelerates the prediction process but, more importantly, can support early-stage inverse design by proposing candidate compositions with targeted elastic properties—subject to the fidelity of the predictive surrogate and downstream verification.

\paragraph{Thermodynamic Stability }
Among the most basic and important thermodynamic properties is the formation energy, which dictates stability for a crystalline solid~\cite{chaib2020effect}. Its prediction is thus of paramount importance in order for high-throughput screening to be performed on new, synthesizable materials. Machine learning models have thus been employed to directly predict it from crystal structures, with the key ingredient being sophisticated feature engineering to account for periodicity in solids~\citep{faber2015crystal}. Subsequent deep learning models demonstrated the feasibility of material property inference from merely elemental composition and symmetry classification~\citep{houchins2024formation}.

A particularly noticeable development in this domain has been the development of structure-agnostic models, which predict properties from stoichiometry alone. Using a message-passing architecture on a dense weighted graph representation of the chemical formula, the Roost framework learns material descriptors in an explicit-structure-free way~\citep{goodall2024pretraining}. This is especially important in the early stage of materials discovery when little to no crystal structural information is known. Self-supervised pre-training methods have been shown to boost the performances of such models even further, with substantial improvements in the low-data regime~\citep{dan2024pretraining}. Furthermore, LLMs have been demonstrated to handle simple chemical formulas and predict complex chemical properties (e.g., formation energy), significantly reducing the computational cost of materials screening.

\paragraph{Thermoelectric and Thermal Properties}
Thermoelectric materials enable the conversion between thermal and electrical energy, making them crucial for waste heat recovery and solid-state cooling technologies. 
However, the lack of large, structured datasets with relevant performance metrics hinders the development of models for high-performance thermoelectric materials~\cite{wang2023critical}.
LLM-based extraction approaches are increasingly used to mitigate this data bottleneck.
Recent work has shown that agentic, LLM-driven workflows autonomously extract the thermoelectric and structural properties from thousands of scientific articles~\citep{ghosh2025automated}. These intelligent agents can parse text, tables, and captions to create large machine-readable datasets that couple performance metrics with structural information such as the crystal class and space group~\citep{korop2025application}. 
This automated curation pipeline can improve downstream regression by expanding training data and improving schema consistency; however, it still primarily optimizes proxy prediction objectives unless coupled to closed-loop validation and decision-making in the discovery loop.

\subsubsection{Classification Tasks\label{sec:3_1_classification}}

Classification is a fundamental task in materials science, referring to the automatic categorization of materials into predefined classes based on their composition, structure, or other characteristic data~\cite{ziletti2018insightful, jha2018elemnet}. Classification is a crucial step in numerous downstream applications, from interpreting experimental data to controlling material quality and accelerating the discovery of new materials~\citep{dunn2020benchmarking}. 
Meanwhile, because LLMs can effectively handle noisy experimental and computational materials data and integrate data inputs from multiple modalities such as text, images, and numerical data~\citep{lei2024materials, schilling2025text}, they are gradually becoming a general solution for classification problems.

\paragraph{Defect Classification and Experimental Automation}
The classification of defects—whether vacancies, dislocations, or grain boundaries—is an important aspect of predicting materials performance and ensuring safety and optimization of manufacturing yield. 
In practice, defect signals may be derived from microscopy~\citep{Paruchuri2024AFM}, in-situ sensors~\citep{Peng2023InSituSLMReview}, or the operational setting ranges from offline labeling to on-the-fly process monitoring~\cite{fang2022process}.
LLM has broad application prospects in this field, serving both static analysis tasks such as defect labeling and integration into closed-loop systems to provide real-time decision support for researchers, thereby accelerating research and development cycles~\citep{bran2024chemcrow, boiko2023emergent}.
At present, most deployments still treat classification as a modular, reactive component; closed-loop autonomy typically requires additional planning, control, and verification modules beyond classification alone.

In additive manufacturing, LLMs fine-tuned on process parameter data can predict defect regimes (i.e. keyholing or lack of fusion) directly from natural language descriptions~\citep{pak2025additivellm}. This enables faster build parameter optimization while significantly reducing reliance on resource-intensive simulations and physical experiments. Similarly, in materials characterization, conversational LLM systems have been developed to support defect classification from Atomic Force Microscopy (AFM) images. These typically involve a two-step pipeline whereby a vision-based model performs an initial defect classification, and then an LLM acts as a conversational assistant. In this latter role, the LLM contextualizes the classification based on queries from users and experimental parameters and provides actionable feedback in natural language, such as suggesting adjustments of scan speed to minimize artifacts~\citep{biswas2025conversational}. 
More broadly, classification becomes substantially more valuable when embedded in an interactive workflow, but the classification objective itself remains a proxy label unless the surrounding system optimizes end-to-end discovery outcomes~\cite{ramos2020interactive}.
This evolution transforms the LLM from a passive classifier into an active participant in the experimental workflow-capable of hierarchical reasoning linking sensor input to corrective measures with reduced reliance on continuous expert supervision~\citep{lei2024materials}. 
Finally, beyond classification, LLM-driven agents are being developed that automate the complete scientific discovery process, from hypothesis generation~\citep{Shahhosseini2025-wn} to experimental design and data analysis~\citep{boiko2023emergent}.

\subsubsection{Advanced Methodologies\label{sec:3_1_3_adv}}

To pursue more robust and generalizable LLMs for materials science, the adoption of deep learning methods has shaped both model architectures and training strategies.

\paragraph{Hybrid GNN-Transformer Architectures}
While GNNs perform well in modeling local atomic environments, their main drawbacks are related to modeling long-range interactions and a well-known issue of over-smoothing in deep architectures~\cite{oono2019graph}. 
Transformers specialize in the opposite: global dependencies with high computational costs for large structures~\cite{vaswani2017attention}. Hybrid GNN-Transformer architectures seek to harness the strengths of both approaches~\cite{yun2019graph}.

These hybrid models generally use a GNN to learn efficiently local structural features from predefined graphs. The node embeddings produced by this step encode structure-aware information and are treated as a sequence of tokens by a Transformer module. This latter component then uses self-attention to infer long-range interactions within the material~\citep{du2024ctgnn}. The hierarchical method therefore overcomes the individual limitations of both architectures by leveraging their combined strengths to afford even more accurate and scalable predictions of material properties~\citep{madani2025accelerating, li2025hybrid}. This synergy thus allows for a rich representation of materials from the atomic to the macroscopic scale.
However, stronger representations alone do not guarantee improved out-of-distribution reliability, motivating uncertainty-aware evaluation and closed-loop validation.

\paragraph{Multi-Task Learning (MTL) for Generalization}
The physical interdependencies between key material properties (such as band gap, elastic modulus, and formation energy) make MTL a very suitable strategy for building generalizable models~\cite{sanyal2018mt}.
By training a single model on these related tasks simultaneously, MTL seeks to discover a unified feature space that inherently encodes the underlying shared representations consistent with physical correlations, resulting in a more robust and generalizable representation and leading to better prediction performance~\citep{caruana1997multitask}.
A practical limitation is negative transfer: performance can degrade when tasks conflict or when task data are imbalanced and heterogeneous, as is common in materials datasets~\citep{aoki2022heterogeneous}.

Architectures such as the Multi-Task Pre-trained Transformer Encoder (MTENCODER) learn general material representations from diverse properties~\citep{prein2023mtencoder}. 
By combining the signals from several auxiliary activities, this method makes data more efficient and helps the models generalize better to new tasks and materials. 
Thus, using MTL in transformer encoders creates strong and transferable embeddings that are an important step toward flexible foundation models for materials science~\cite{shoghi2023molecules}.
From a pipeline-centric perspective, MTL performs internal credit sharing across proxy tasks; aligning task selection and weighting with discovery-level objectives remains an open systems challenge~\cite{ruder2017overview}.

\paragraph{Uncertainty Quantification for Trustworthy AI}
To make AI-driven materials discovery effective, we need a system capable of accurately measuring prediction uncertainty. This measurement of uncertainty is very important for building trust in AI systems, directing active learning strategies, and reliably finding samples that are outside of the distribution~\citep{tavazza2021uncertainty}.

In closed-loop discovery, uncertainty is also a decision variable (e.g., acquisition or stopping).
There are two main kinds of uncertainty quantification methods: aleatoric uncertainty, which comes from noise in the data, and epistemic uncertainty, arising from limited data coverage and model misspecification~\cite{kendall2017uncertainties}.
There are different ways to figure out prediction intervals for material properties. Some of these are probabilistic methods like Gaussian processes, while others are quantile loss functions and methods that directly model prediction error~\citep{tavazza2021uncertainty}. 
Having good estimates of epistemic uncertainty is very helpful because they show parts of chemical space where the model is not very certain. 
This directly benefits targeted data collection, which means that future experiments or simulations could be guided by an active learning loop.
This makes the process of discovering new findings faster and uses fewer resources~\citep{jacobs2023role}. 
In materials settings, uncertainty estimation is further complicated by multi-fidelity data (simulation vs. experiment), distribution shift across conditions, and the compositional long tail—precisely where naïve confidence estimates can fail.
More and more studies are now trying to determine how much uncertainty there is in the outputs of LLMs. 
This is a necessary step for using them reliably in important scientific work~\citep{xiong2024efficient}.
Finally, while uncertainty quantification can make proxy-task predictors more reliable, end-to-end discovery requires propagating uncertainty through multi-step planning and experimentation decisions.

\subsection{Mining\label{sec:3_2_mining}}
\subsubsection{Information Extraction\label{sec:3_2_1_ie}}

Information extraction in materials science focuses on converting experimental descriptions and performance data from publications into structured formats suitable for data-driven research. 
As a pipeline-critical bottleneck, extraction determines what becomes trainable downstream (datasets, benchmarks, and ultimately the objectives that models optimize).
Early works used rule-based text mining systems such as ChemDataExtractor~\cite{swain2016chemdataextractor}, which automatically produced databases of synthesis parameters~\cite{kim2017sdata_oxide}, battery compositions~\cite{cde_battery_2020_sdata}, thermoelectric properties~\cite{sierepeklis2022thermo_sdata}, and mechanical data~\cite{cde_stressstrain_2024_sdata} from journal article. These approaches provided reproducible extraction pipelines but were constrained by fixed grammars and limited adaptability to variations in scientific writing.

To address these limitations, subsequent research introduced statistical and representation-learning approaches. Embedding models trained on large-scale materials corpora captured latent chemical and physical relations through co-occurrence patterns~\cite{tshitoyan2019unsupervised}, while domain-specific advances in named entity recognition and normalization improved terminology consistency \cite{jcim_ner_norm_2019}. Building on these foundations, transformer-based models such as MatSciBERT \citep{gupta2022matscibert} and MechBERT \citep{kumar2025mechbert} provided contextualized embeddings that greatly improved extraction robustness and accuracy across diverse publication sources.

Research on LLMs for materials science, utilizing prompt engineering and fine-tuning, has expanded the limits of extraction system. ChatExtract~\citep{polak2024extracting} builds a conversational system with engineered prompts and follow-up questions to identify evidence sentences, extract structured values, and self-verify, reaching high precision and recall suitable for database curation. 
Similarly, MaTableGPT \citep{adv_sci_matablegpt_2025}, an instruction-tuned and table-aware model, jointly reasons over narrative text and tabular data for quantitative extraction. 
\citet{ai4mat2024_torres_mof_llm_extract} presented an automated pipeline for extracting synthesis procedures for reticular materials from PDFs using prompt-based paragraph classification. \citet{ai4mat2025_wang_llm_decision} also explored LLM-augmented decision programs that link extracted synthesis data to design workflows. 
At the same time, others combine extraction and verification mechanisms to autonomously construct structural datasets from full-text literature \citep{gupta2024data, ghosh2025automated, schilling2025text}. 
Further, ~\citet{ansari2024agent} developed a chemist AI agent that extracts and structures materials science data with chain-of-verification tool to create datasets.
Collectively, these efforts suggest a shift from one-shot parsing toward agentic extraction pipelines, and yet extraction still functions largely as a reactive module unless it is coupled to discovery-level objectives and outcome-grounded evaluation.

\subsubsection{Knowledge Graph\label{sec:3_2_2_kg}}
As extraction quality improved, researchers began to integrate extracted entities and relations into structured knowledge representations. Hybrid and semi-supervised methods complement purely model-based extraction by building structured knowledge representations. Hybrid methods combine GNNs, structured ontology models and LLMs to extend knowledge graphs and interpret predicted relations~\cite{ivanisenko2024accurate}.
Semi-supervised pipelines have been applied to identify synthesis–processing–property relations in alloy systems \citep{npj_alloy_semisup_2023}. Machine learning combined with text mining has enabled the reconstruction of magnetic and superconducting phase diagrams from literature \citep{npj_phase_diagrams_2020}. These extracted relations feed into knowledge graphs such as MatKG and terminology-based networks \citep{sdata_matkg_2024,sdata_termkg_2024}, which integrate entities, parameters, and outcomes into a structured representation for downstream reasoning across the composition–structure–property space. 
\citet{ai4mat2025_prein_retrorankin} proposed a ranking-based approach to inorganic synthesis planning that learns precursor relationships in a latent space, providing new insights for reaction network construction.
A key limitation is that KG coverage and correctness remain bounded by extraction noise and ontology choices, and thus errors could propagate into downstream reasoning and planning~\cite{paulheim2016knowledge}.

\subsubsection{Database Automation and Annotation\label{sec:3_2_3_db}}
Large automatically generated corpora \citep{sdata_auto_corpus_2022} enabled standardized benchmarks for entity and relation extraction. Together, these efforts shifted the extraction paradigm from rigid rules to learned semantic representations. Frameworks like MatNexus \citep{softwarex_matnexus_2024} and MatScIE \citep{cms_matscie_2021} provide modular pipelines to aggregate extraction outputs and align them with ontologies. CALMS \citep{npj_calms_2024} proposed linking LLM-based extraction with instrument control and database queries. Multimodal extraction combining text and tables \citep{cms_text_table_2023} and literature-based reconstruction of the materials tetrahedron \citep{digitaldisc_tetrahedron_2024} illustrated progress toward unified modeling of composition, structure, process, and property. 
From a pipeline-centric perspective, automated corpora and benchmarks can accelerate iteration, but they can also freeze proxy objectives when benchmark definitions drift from discovery outcome-grounded evaluation and continual dataset expansion and revision.
In practice, rule-free and model-agnostic curation workflows now turn full texts into mid-sized, high-precision databases with minimal bespoke coding and near-perfect precision–recall trade-offs, making them ideal for rapid property-focused datasets.

\subsection{Generation\label{sec:3_3_generation}}
The generative paradigm has also profoundly shaped AI for materials science. LLMs, in particular, can be employed for designing novel materials and synthesis methods, expanding materials R\&D beyond data analysis into proactive innovation. Key tasks include structure generation, inverse design targeting desired properties, and synthesis route generation.

\subsubsection{Structure Generation\label{sec:3_3_1_stru}}
The task of structure generation in AI4Materials spans multiple scales and materials classes, aiming to propose new candidate structures that are valid, stable, and potentially synthesizable~\cite{zeni2024mattergengenerativemodelinorganic}.
In practice, these criteria are often operationalized using surrogates (e.g., predicted formation energy), which can diverge from real synthesis feasibility—motivating the pipeline-centric emphasis on verification and closed-loop feedback.
While early research focused on crystalline solids—where periodicity, symmetry, and lattice constraints dominate—the field now extends to polymers, porous frameworks (MOFs~\citep{MOFFUSION2025NatComms}/COFs~\citep{Kim2020InverseDesign}), amorphous phases such as glasses and electrolytes, microstructures at the mesoscale, and even architected metamaterials.
In these fields, materials can be represented as matrices, voxels, or graphs, and generative models~\citep{Metni2025GenerativeModels} are used to generate realistic atomic structures.
For discovery, the key differentiators are less the model family than whether generation is coupled to evaluation in the loop.

For crystalline materials, a complementary research approach has emerged that represents crystal structures as text sequences and generates them using language models. 
This language-based paradigm treats structural information as a sequence of tokens, which makes it possible to generate lattices and compositions in an autoregressive manner. Following this idea, CrystalLLM\citep{antunes2024crystalstructuregenerationautoregressive} further highlighted the potential of sequential generation by applying autoregressive language modeling to produce CIF file sequences directly. \citet{gruver2024finetuned} further demonstrated that fine-tuned LLaMA-2 models can directly generate stable inorganic crystal structures purely as text, while achieving nearly double the generation speed that of diffusion-based baselines (i.e. CDVAE\citep{xie2022cdvae}). More recently, MATLLMSEARCH~\citep{gan2025large} showed that pre-trained LLMs can act as agentic generators of crystal structures without additional fine-tuning. Their system autonomously explores and optimizes materials by combining a pre-trained LLaMA-3.1 model with evolutionary search and physical evaluation. This shows a shift in regarding LLMs as reasoning agents that can iteratively search, evaluate, and improve candidate structures.
Collectively, these works mark a transition from generative sampling to search-with-verification, which is loser to our pipeline-centric view. 
Unfortunately, most evaluations still rely primarily on simulated or surrogate validators.

\subsubsection{Inverse Design\label{sec:3_3_2_id}}  
In materials generation using generative models, the objective is typically to produce new materials that follow the distribution of the training data. 
However, in practical materials discovery, there is often a demand to identify materials that exhibit specific desired properties. 
MatAgent~\citep{takahara2025acceleratedinorganicmaterialsdesign}, proposes an agentic materials generation framework combining an LLM, a diffusion model, and a property prediction model. In this framework, the LLM proposes candidate compositions and iteratively refines them based on feedback from the property predictor, enabling the autonomous generation of materials with desired properties. RL has also entered this space: to inverse inorganic oxide materials design to target promising compounds using specified property and synthesis objectives \citep{pan2022deepreinforcementlearninginverse}. 
Because RL naturally supports long-horizon objectives, careful reward and constraint design is essential to avoid optimizing proxy signals that do not translate to real discovery outcomes.
Deep dreaming solves the inverse design challenge for metal–organic frameworks by directly optimizing string-based representations of MOF linkers within a unified neural network to generate novel, synthesizable structures with targeted properties~\citep{cleeton2025inverse}. 
MatDesINNe is a framework based on invertible neural networks (INNs) that learns both forward and reverse mappings between design parameters and target properties, enabling generation of candidate materials for specified property goals (e.g., band gaps in 2D materials) \citep{fung2021inversedesigntwodimensionalmaterials}. 
PolyTAO (Transformer-Assisted Oriented model) is a generative pre-trained model for polymers: given a vector of target polymer properties, it generates polymer repeat unit SMILES with very high chemical validity and accurately matches desired property values in semi-template or even template-free settings \citep{qiu2024demand}.   
These approaches share a common dependency on surrogate property models and feasibility constraints, motivating closed-loop verification later in the discovery pipeline.

\subsubsection{Synthesis Route Generation\label{sec:3_3_3_srg}}
Synthesis route generation refers to constructing feasible pathways to synthesize a target material. In materials science, it covers various aspects, such as choosing suitable precursors, suggesting reaction paths, and making predictions about key reaction conditions like time, temperature, and reaction environment~\cite{zhang2024evolutionary}.
In practice, synthesis is often the dominant bottleneck: feasibility and process constraints are what separate generated candidates from experimentally validated materials.
Graph-based methodologies were first applied to depict solid-state processes and employed pathfinding to ascertain viable synthesis pathways \citep{mcdermott2021graph}.  
Researchers have also used machine learning to directly predict experimental conditions from data in the literature. This has shown links between the stability of precursors and the temperatures or periods of synthesis\citep{huo2022machine}. 
In addition to reactions, models can suggest sets of precursors based on how similar the materials are, which makes recipes for inorganic synthesis more accurate\citep{he2023precursor}.
At the system level, autonomous laboratories are already combining AI planning with robotics and active learning. 
For example, A-Lab can design, run, and refine synthesis strategies for new inorganic compounds iteratively again\citep{szymanski2023autonomous}. 
What makes such platforms non-reactive is closed-loop execution, where synthesis outcomes update models and trigger replanning, rather than ending at a one-shot proposal.

Synthesis planning is a complex task requiring the collaborative optimization of advanced algorithms such as deep learning, RL, and uncertainty-aware search~\cite{yuan2024active}.
The aforementioned approaches reveal that synthesis planning is evolving toward a data-driven closed-loop system.
The systems must not only generate synthesis paths tailored to specific requirements but also perform real-time updates based on synthesis outcomes~\cite{song2025aot}.
These loops can supply discovery-level feedback signals that support end-to-end credit assignment across upstream components.

\subsection{Optimization and Verification\label{sec:3_4_optimization}}
\subsubsection{Materials Discovery Process Optimization\label{sec:3_4_1_mdpo}}

As an instance of the optimization lab, a self-driving platform, Ada~\citep{macleod2022self}, has been developed; an AI-driven lab for the synthesis of palladium thin films that uses a Bayesian optimization algorithm to autonomously optimize the process in a closed loop where objectives are measured on experimental outcomes.
By mapping the conductivity-temperature trade-off, the AI model identifies the optimal low-temperature, high-conductivity formulations, which were later successfully validated in a scalable manufacturing process. Later, ~\citet{chitturi2024targeted} demonstrated an autonomous material search platform by developing a multi-property version of Bayesian algorithm execution (BAX), which transforms the user-defined targets into acquisition functions to guide the autonomous targeted materials discovery. 
Reported evaluations emphasize efficiency in locating target materials; for discovery-oriented settings, these should be interpreted through outcome-level metrics (e.g., experimental sample efficiency), not benchmark-style scores alone.
Benchmarks showed that variants of BAX outperform the strong baseline methods in efficiently locating target materials, but only in low-dimensional settings~\citep{Wang2016REMBO}.

SPINBOT \citep{zhang2023optimizing}, an autonomous platform employing Bayesian optimization, optimizes perovskite film fabrication based on photoluminescence feedback. Similarly, \citet{chang2020efficient} developed an Autonomous Research System (ARES) utilized Bayesian optimization to accelerate carbon nanotube synthesis, achieving an 8-fold increase in the growth rate of carbon nanotubes, demonstrating the effectiveness of closed-loop optimization. Besides Bayesian optimization, RLHF is also used in M$^{4}$olGen \cite{li2026m4olgenmultiagentmultistagemolecular} to achieve multi-objective optimization on property-constrained molecular generation.

\subsubsection{Simulation and AI-based Verification\label{sec:3_4_2_simulation}}
\citet{fare2022multi} proposed a method by leveraging the multi-output Gaussian process that dynamically fuses information from sources of varying accuracy and cost (e.g., simulations and experiments). The optimization task is driven by Targeted Variance Reduction with Expected Improvement (TVR-EI), a multi-fidelity Bayesian algorithm, verified by performing computational simulations. 
More broadly, verification in materials discovery is inherently multi-fidelity (approximate vs accurate), so pipeline-centric systems must allocate verification budgets adaptively.
\citet{merchant2023scaling} demonstrated Graph Networks for Materials Exploration (GNOME) for large-scale materials discovery by combining GNNs with Density Functional Theory (DFT) simulations. 

For the autonomous investigation of the mechanism of catalytic reactions, \citet{lan2024enabling} applied deep RL methods by developing a high-throughput framework coupled with first-principles DFT simulations. 
More recently, \citet{wen2025cartesian} showcased the integration of Cartesian Atomic Moment Potential (CAMP), a machine-learned interatomic potential (MLIP) built entirely in Cartesian space, in closed-loop simulations, as the next generation MLIP primitive. They have demonstrated that CAMP offers systematic improvements through utilizing physically motivated atomic moment tensors and their tensor products within a graph neural network framework to capture local atomic environments and higher-order interactions. 
From a pipeline-centric view, improved MLIPs matter because they reduce verification cost and expand the feasible horizon of simulation-driven loops, making longer-horizon, agentic planning more practical.

\subsubsection{Agent-based Closed-loop Labs\label{sec:3_4_3_closed_lab}}
Closed-loop labs accelerate research through the integration of AI with automated and self-driving experiments~\citep{kulichenko2024data, abolhasani2023rise, seifrid2022autonomous, angello2024closed, tom2024self}. The ``mobile robotic chemist'' was presented by \citet{burger2020mobile}, employing a free-roaming robot, as an autonomous agent, to automate the researcher's tasks. The robot was tasked with searching for an improved photocatalyst, driven by a batched Bayesian search algorithm exploring a ten-variable experimental space, and Gaussian process regression was used to build a predictive model. The ``Artificial Chemist'', a self-driven agent, was developed by \citet{epps2020artificial} by coupling an autonomous microfluidic flow reactor with machine learning to optimize quantum dot synthesis. 
Later \citet{sadeghi2024autonomous} developed a self-driving fluidic lab by integrating microfluidics, automation, and Bayesian optimization for the autonomous synthesis of lead-free perovskite nanocrystals. 
\citet{Kusne2020NatCommun} presented CAMEO, closed-loop autonomous materials exploration and optimization, an autonomous agent designed to discover new inorganic materials, driven by Bayesian active learning to explore the complex relationship among the composition, structural phase, and properties of the material. The ``A-Lab'' was developed by \citet{szymanski2023autonomous}, an autonomous platform for the synthesis of novel solid-state inorganic materials. The A-Lab platform employed natural language processing models trained on literature to propose initial recipes, and optimization was carried out by ARROWS~\citep{szymanski2023dynamic}, an active learning algorithm that learns from previous experimental outcomes. Upon failure of synthesis, (ARROWS) uses ab initio computed reaction energies to predict and optimize alternative reaction pathways.
More recently, \citet{omidvar2024accelerated} presented an agent-based closed-loop platform for the synthesis of perovskite solid solutions by combining ML-guided composition prediction, automated solid-state synthesis, and high-throughput characterization.

LLMs have broadened self-directed materials exploration. 
In the realm of the application of LLMs in chemical synthesis, Coscientist, an autonomous chemical research framework, has been developed by employing GPT-4-based agents to orchestrate chemical experiments~\citep{boiko2023autonomous}. 
Additionally, in materials characterization, LLMs are currently employed not only for experimental design but also for instrument control and automation via code generation. 
For example, one implementation demonstrated that an LLM could independently write executable scripts (e.g., Python and LabVIEW) to control laboratory instruments from natural language prompts, showcasing genuine hardware control with minimal human involvement~\citep{liu2024synergizing}. 
In recent times, LLMs have been used to achieve full laboratory autonomy by bringing together experiment design, instrument operation, data collection, and analysis into one system.
More recently, LLM-centered systems have been explored for tighter end-to-end loops that combine experiment design, instrument operation, data collection, and analysis.
\citet{xie2025toward} described a comprehensive system in which the LLM not only produced control code but also analyzed experimental outcomes and enhanced subsequent measurement circumstances, thereby establishing a closed-loop connection between thinking and execution. 
These research efforts clearly show that we are moving toward experimental autonomy based on natural language processing. In this new way, LLMs are not just instruments for communication; they are also agents for coding, control, and reasoning that connect human purpose with machine execution.

To integrate human experts with machine learning, \citet{adams2024human} developed a human-in-the-loop Bayesian framework where expert feedback is incorporated as priors to guide automated phase mapping experiments.
BOARS addressed the rigidity of predefined targets in autonomous experiments by developing a human-in-the-loop workflow \citep{biswas2024dynamic}. 
The optimization strategy allows a human operator to dynamically define and refine the scientific objective during the experiment by upvoting or downvoting real-time spectral measurements, successfully locating the human-desired optimum in significantly less time compared to manual optimization.

\subsection{Data and Knowledge\label{sec:3_5_reactive_challenges_data}}

One of the long-standing challenges in applying AI to materials science is the scarcity and heterogeneity of material data and knowledge.
This becomes more challenging for an end-to-end agentic LLM for materials science, especially considering the costs associated with evaluating the outputs in the real-world, never-ending experimental environment. 

\subsubsection{Data scarcity and heterogeneity\label{sec:3_5_1_data_scarcity}}
The advancement of data-driven materials science is characterized by the collaborative endeavors of all stakeholders to address the challenges of data scarcity and data heterogeneity~\cite{xu2023small, medina2022accelerating}.
To solve this problem, systems like MatMiner~\citep{ward2018matminer} and Robocrystallographer~\citep{ganose2019robocrystallographer} give us the tools we need to turn different types of materials into standardized numerical descriptors that ML can use.  
MatMiner has a full library of features that cover a wide range of attributes. Robocrystallographer, on the other hand, is made to create interpretable features from crystallographic data.

OPTIMADE~\citep{andersen2021optimade} provides a unified API enabling researchers to query diverse materials databases simultaneously, thereby partially addressing data heterogeneity issues. 
\citet{ghiringhelli2023shared} proposes a hierarchical metadata approach,FAIR, which stands for Findable, Accessible, Interoperable, and Reusable, advocating for a unified and streamlined philosophy for computational and experimental data management.

The LeMatTraj project~\citep{ramlaoui2025lemat} directly solves the problem of not having enough data by combining nearly 120 million high-quality atomic configurations from the primary repository into a single dataset. This is in addition to the infrastructure standards. 
This huge collection has the high-quality, standardized data that is needed to develop strong machine learning interatomic potentials. 

\paragraph{Data Augmentation\label{sec:data_augmentation}}

Data augmentation techniques are crucial for mitigating data scarcity and heterogeneity in materials modeling, evolving from basic perturbation strategies to complex physics-informed generative methods.
Early studies, such as OC20~\citep{chanussot2021open}, supplemented DFT datasets through random structural perturbations, short-timescale molecular dynamics simulations, and electronic structure analysis.
 It then slowly changed into physics-informed augmentation strategies: 
\citet{szymanski2024integrated} included experimental errors (lattice strain, crystal texture) in the XRD pattern generation process, thus linking simulations to experiments.
\citet{gibson2022data} proposed a physically motivated, computationally cheap perturbation technique that augments training data to improve the accuracy of unrelaxed structure prediction by 66\% and directly addressing the domain mismatch problem in crystal structure prediction.
Based on this idea, \citet{dinic2023strain} presented an ML model capable of predicting the crystal energy response to global strain by using available elasticity data to augment the dataset, which greatly improved energy forecasts for structures that were not perfectly straight.  
In addition to physics-motivated methods, generative techniques that produce high-quality in-silico data to augment the original dataset have developed.\citet{chung2024imbalanced} proposed a GAN-based data synthesis method to produce spectral samples to address the class imbalance problem, thereby enhancing the performance of the ML model trained on the augmented dataset. 
MatWheel~\citep{li2025matwheel} introduced a framework that trains the materials property prediction model using the synthetic data generated by the conditional generative model, showing its potential in addressing extreme data-scarce scenarios in materials science.
% Because synthetic data can amplify existing model biases, it is most reliable when combined with uncertainty-aware filtering and, where feasible, experimental validation.

\paragraph{Multi-fidelity learning}
Due to the trade-off between cost and accuracy, materials science involves measurements and simulation data with varying degrees of precision. To enable the fusion learning of existing data with different levels of accuracy, researchers have proposed multi-fidelity learning. Its objective is to maximize the utilization of existing data for high-accuracy model predictions while minimizing additional computational and human resource costs.
\citet{pilania2017multi} introduced this approach using co-kriging to fuse low-cost, low-accuracy DFT calculations with sparse high-fidelity data, achieving accurate electronic band gap predictions at significantly reduced computational cost.
Further, MODNet~\citep{debreuck2022accurate} replaced Gaussian process regression with DNN while learning about the differences between high- and low-quality values, which shows significant improvement in the results compared to learning on the sole high-quality experimental data.
In contrast, MD-HIT~\citep{li2024md} addresses the redundancy problem inherent in multi-source datasets, employing greedy incremental algorithms to filter overlapping samples and prevent overestimation of performance, thereby enabling more realistic evaluation and improved generalization to out-of-distribution materials.

\paragraph{Few-shot Learning}
Few-shot learning, which is training a model on a small number of labeled examples, can help models quickly learn new tasks, especially when there is not a lot of good training data available.
Meta-MGNN~\citep{guo2021few} was the first works creating meta-learning frameworks for predicting molecular characteristics. This framework allows models to quickly adapt to new attributes with training on a few instances, and also use self-supervised learning from molecular structures to make up for the lack of labeled data. AttFPGNN-MAML~\citep{qian2024meta} built on this by combining hybrid molecular representations fingerprints and graph neural networks with ProtoMAML meta-learning. This made it easier to adapt to new prediction tasks. Most recently, MolFeSCue~\citep{zhang2024molfescue} moved the field forward by adding dynamic contrastive loss functions that let pre-trained models learn useful representations from very unbalanced and small amounts of training data. It addresses the challenge of data scarcity in drug discovery and materials research.

\subsubsection{Knowledge integration\label{sec:3_5_2_knowledge_integration}}
Knowledge integration in AI4MS refers to using relevant knowledge from materials science to guide the design and training of AI models, thereby improving the accuracy, interpretability, and generalization ability of the models.

\paragraph{Physics-informed Neural Networks(PINNs)} 
Physics-informed neural networks (PINNs) are a type of general-purpose function approximator that can embed knowledge of any physical laws governing a given dataset into the learning process.
\citet{chen2020physics} employs PINNs for the solution of inverse scattering problems that retrieves the effective permittivity parameters of a number of finite-size scattering systems in photonic metamaterials and nanooptics.
\citet{fang2019deep} applied deep physics-informed neural networks for electromagnetic metamaterial design, which can be utilized in dealing with various practical problems such as cloaking, rotators, concentrators, etc.
\citet{zhang2022analyses} employ PINNs for defect characterization in mechanical materials, using equilibrium PDEs and constitutive laws to identify internal voids/inclusions from boundary displacement measurements. 

\paragraph{Structure-based Knowledge Integration}

ElemNet~\citep{jha2018elemnet} and CrabNet~\citep{wang2021compositionally} only use the elemental composition of the material as input. They automatically learn chemical relationships and composition-property correlations between elements based on chemical relationships in the periodic table, without requiring any manually designed descriptors.
In addition, GNNs such as MEGNet~\citep{chen2019graph}, ALIGNN~\citep{Choudhary2021Atomistic}, ROOST~\citep{goodall2020predicting}, and GnoME~\citep{merchant2023scaling} model molecules and crystals by treating atoms as nodes, bonds as edges, and selectively adding global features such as temperature or pressure. This approach allows deep learning algorithms to make very accurate predictions of features such as formation energy, band gap, and elastic modulus.
Based on this general modeling, many new studies have incorporated additional prior knowledge of material structure features by introducing new encoding modules or learning objectives to integrate structural knowledge:
M3GNet~\citep{m3gnet2022} introduced a many-body computation module to calculate the three-body and many-body interactions;
NequIP~\citep{batzner2022e3} proposed E(3)-equivariant neural networks using geometric tensors and equivariant convolutions based on tensor products with spherical harmonics, which introduces feature encoding of the rotational and translational symmetries of the material structure. 
Allegro~\citep{musaelian2023learning} removes atom-centered message forwarding while maintaining equivariance by employing learned representations in iterated tensor products.

\paragraph{Ontology and Knowledge Graph Integration} 
As discussed in Section~\ref{sec:3_2_mining}, extensive knowledge bases and knowledge graphs have been constructed from various sources, including academic articles and experimental analyses. This organized knowledge in databases, covering multiple kinds of knowledge in materials science, can then be used to enhance model performance in material property prediction tasks.
\citet{zhao2021knowledge} is specifically designed to address the challenges of data heterogeneity in materials science. It uses knowledge graphs as a semantic integration center and combines ontology mapping and automatic annotation to extract information from academic articles.
Both KCL~\citep{fang2022molecular}, KANO~\citep{fang2023knowledge}, ESNet~\citep{huang2024material} construct an element-centric knowledge graph capturing intrinsic physical and chemical attributes of elements and leverages it in a multimodal or contrastive learning framework to unify structural features with semantic chemical knowledge for improved molecular and materials property prediction.
Recent innovations combine LLM with knowledge integration: MOFs-KG~\citep{an2023knowledge} createed the first comprehensive KGQA benchmark specifically for materials science (MOFs) with 644 complex natural language questions, demonstrating ChatGPT's systematic application to domain-specific knowledge graphs.

\paragraph{Pre-trained Model}
Pre-trained models have been demonstrated to be effective in natural language processing, especially in data scarcity scenarios. Through training with limited data, pre-trained models can be transferred to a specific task while showing strong performance. In materials science, pre-trained model are also widely used. 
Early work, like Mat2Vec~\citep{tshitoyan2019unsupervised}, learned word embeddings in materials literature without human supervision, showing a surprising capability to identify semantics in materials science text.
This led to the development of material-specific language models such as MatSciBERT~\citep{gupta2022matscibert} and MaterialBERT~\citep{shetty2023general}. They collect a large corpus of materials science for continual training without changing the BERT architecture.
Compared to vanilla BERT, they show superior performance in identifying material-related property data and phrases in the literature. This conclusion contributes to the emergence of various BERT variations that were particular to certain materials subfields, such as BatteryBERT~\citep{huang2022batterybert} for battery-related material, OpticalBERT~\citep{zhao2023opticalbert} for optical materials, PolyBERT~\citep{kuenneth2023polybert} for polymers materials, and SteelBERT~\citep{tian2025steel} for metallurgy materials.
UniMat~\citep{ock2024unimat} propelled the field forward by merging multiple modes (atomic structure, XRD patterns, and composition) into strong joint embeddings. This made it easier to find novel materials. SmileyLlama~\citep{cavanagh2025smileyllamamodifyinglargelanguage} is further eolved with Direct Preference Optimization(DPO) to optimize the structure generation.
Recent ideas deal with structural limitations: \citet{huang2024pretraining} proposed a self-supervised and multimodal pre-training technique that can predict performance based only on stoichiometry, without explicit understanding of the atomic structure. 
MELT\citep{kim2024melt} was the first to employ curriculum-based continuous pre-training, which used semantic knowledge graphs to convert models from broad-domain knowledge to material-specific concepts in a systematic way.

\subsubsection{Multimodality\label{sec:3_5_3_multimodality_challenges}}

Recent advances in applying artificial intelligence to materials research have demonstrated the benefits of domain-specific language models, graph-based predictors, and image analysis frameworks~\cite{shetty2023general, musaelian2023learning, matvqa2025}. 
To effectively discover novel materials, agentic LLMs must process and understand multimodal data at each stage of the pipeline, not only as more inputs, but as a prerequisite for closing the loop, where each stage and component produces distinct modalities that must be jointly interpreted for decision-making~\cite{Yao2022, wang2023voyager, plaat2025agentic}. 
Most existing improvements are limited to single-model scenarios, where algorithms are designed for only a single type of data.  
However, scientific research often uses multiple types of input for representation, such as text reports, structural graphs, microscopy images, and spectroscopic signals. Each representation reveals the material's properties from a different perspective. 
Integrating different types of data into a single analytical framework is challenging due to methodological and infrastructure issues.
This section will explore these issues, first introducing the characteristics of multimodal scientific data, and then gradually delving into the various methods designed to integrate the data, with success defined in terms of cross-modal generalization, more reliable mechanistic inference, and improved closed-loop decision-making.

\paragraph{Multimodal Scientific Data}
Materials science research increasingly relies on multiple data sources, each revealing different aspects of material properties, thus requiring tailored data analysis methods.
Using domain-specific language models, textual information from publications, patents, and experimental protocols has been put into order. 
MatSciBERT \citep{gupta2022matscibert} showed that it is possible to pull out entities and relationships from scientific literature. 
The MatSci-NLP benchmark \citep{matscinlp2023} set up standard tasks for synthesis action extraction, which made it possible to systematically compare and make further analysis on current methods.

Image-based modalities provide further information at both mesoscopic and microscopic scales. There are big improvements in deep learning for microscopy, but a complete study \citep{microscopyreview2023} points out that some problems remains, such as failing to do 3D reconstruction, high reliance on synthetic data, and lacking enough annotated dataset.  
Another approach is through spectroscopic and diffraction data.  
Interpretable XRD analysis \citep{xrd2025} shows that multi-phase quantification remain difficult, especially in the absence of simulation-based guidance. DiffractGPT \citep{diffractgpt2025} analyzes diffraction profiles in a novel way, treating them as sequence data, enabling autoregressive modeling methods to directly infer atomic structures.

Recent workshops have enriched the research environment by providing multimodal datasets and benchmarks. The TDCM25 dataset~\citep{polat2025tdcm} gathers crystalline materials that change with temperature, along with aligned structural, image, and textual characteristics, which helps with research on phase transitions. 
Similarly, ~\citet{takeda2023multimodal} showed a multimodal fundamental dataset that integrates SELFIES molecular representations, DFT-calculated properties, and spectral characteristics to improve the accuracy of property predictions. 
NanoMINER \citep{odobesku2025nanominer} utilized language models to extract structured knowledge on nanomaterials, connecting unstructured literature with structured graphs.  
These improvements show that multimodal resources are becoming more comprehensive, but they also show that there is still no common metadata or ontologies that can be used with each other.  
Because of this, it is still hard to reproduce results across different modalities. 

Beyond crystalline and bulk materials, molecular discovery introduces additional representational challenges that require foundation models to be ‘multilingual’ across structural, physicochemical, and knowledge-based modalities.
To accelerate scientific discovery, a foundation model must become ``multilingual,'' able to interpret the vast modalities of scientific data beyond text. This includes 2D~\citep{molecule2022} and 3D~\citep{li2024geometryinformedtokenizationmolecules} structures described using geometric graphs, physicochemical properties (e.g., chirality, electrostatic potentials, and hydrophobicity), and dynamic data from molecular and quantum dynamics (MD/QD). It must also incorporate knowledge from literature~\citep{mCLM2025} and knowledge bases~\citep{BiomedicalLM2023}, such as images, charts~\cite{chart2023}, and reaction pathways~\citep{molecule2022}. 

Some recent work proposed joint molecule and language learning~\citep{MolT5} and applied it to power conversion efficiency for organic photovoltaic (OPV) devices~\citep{GLaD2024}. However, most molecule language models tokenize at the level of atoms and/or they do not infuse synthesis considerations up front, which precludes matching tokens with desired functions and/or leads to AI-generated new molecular targets that are not readily synthesizable. To address these challenges, the recent highly innovative, function- and synthesis-aware modular Chemical Language Model (mCLM)~\citep{mCLM2025} proposed a unified multimodal architecture and a suite of specialized encoders to tokenize each modality into a biologically and chemically meaningful language that can be reasoned over. 
mCLM incorporates function- and synthesis-related knowledge into the molecule discovery process a priori. mCLM generates new data on demand in the forward direction by an AI guided closed loop, where a human scientist simply inputs a few natural language phrases or sentences describing the desired functions of a molecule and receives the molecular structure with the best synthesizable functional modules in return.

However, some additional challenges remain. The self-attention mechanism in the Transformer operates by allowing each token to aggregate information from other tokens based on semantic relevance~\cite{Vaswani2017}. 
This formulation is rooted in natural language processing, where word embeddings capture semantic associations and attention primarily aggregates unary information across tokens. 
However, such a design may not align well with materials discovery in the physical world, where governing mechanisms are often determined by explicit pairwise or relational interactions rather than semantic similarity~\cite{gilmer2017neural, battaglia2018relational}. 
For example, in organic photovoltaic (OPV) materials, key performance-determining factors—such as donor–acceptor energy level alignment and balanced charge transport—are jointly determined by interacting molecular pairs~\cite{scharber2006design, brabec2003organic}. 
This mismatch suggests that future research should explore attention mechanisms that explicitly encode binary or relational “mutual influence” between material entities, introducing inductive biases more consistent with physical interaction mechanisms~\cite{battaglia2018relational, gilmer2017neural}. 
Besides, while unified embedding-based multimodal models often improve predictive accuracy, they frequently provide limited mechanistic insight, raising concerns that learned representations may reflect correlations rather than causal or physically grounded relationships~\cite{schutt2017quantum}.

\paragraph{Multimodal Fusion}
Recent methodological advancements have concentrated on developing unified representation frameworks that integrate multiple modalities into a single representation space. MultiMat \citep{moro2025multimodal} presented a self-supervised foundation model that simultaneously encodes text, structural graphs, microscope pictures, and spectroscopic profiles, thereby augmenting transfer learning for subsequent property prediction.  
MatBind \citep{mirza2025matbind} built upon this by using contrastive learning aims to align density-of-states, crystal structures, and textual descriptions, achieving strong cross-modal retrieval.  
MatMCL \citep{matmcl2025} also enhanced the robustness of the model when dealing with data that only has some modalities, proving that multimodal information is redundant and complementary, and that the model can maintain prediction accuracy even when one or more modal information is missing.

Other solutions change the way language models work, enabling them to handle non-textual data.
For example, DiffractGPT \citep{diffractgpt2025} uses sequential modeling approaches to analyze diffraction spectra. This shows that methods originally designed for natural language can also be applied to crystallography. Hybrid systems that combine (GNNs, such as  M3GNet \citep{m3gnet2022}) with LLMs further demonstrate how atomic representations can be linked to contextual knowledge in text. 
There are also new efforts to set benchmarks. MatVQA \citep{matvqa2025} establishes multimodal question answering using figures and text, whereas MaCBench \citep{macbench2024} consolidates band structures, microscope images, and literature, facilitating a systematic assessment of general-purpose multimodal models in scientific endeavors.  
Fusion models like MATMMFuse \citep{pai2025reliability} use multi-head attention to combine crystal graph embeddings with SciBERT textual encoders. This consistently makes it easier to predict formation energies and electrical characteristics.  
However, most current multimodal benchmarks emphasize recognition and cross-modal alignment and only partially evaluate multi-step scientific reasoning or closed-loop decision-making—the regimes most relevant to autonomous discovery.

These strategies substantially help bring together different signals, but there are still certain problems that need to be fixed.  
Unified embeddings can improve forecasts, but they don't help us understand how things work very well.  
This raises the possibility that the captured relationship might show correlations instead of causal ties.  
Using high-dimensional modalities like microscope images increases computational and data-engineering costs, which makes them harder to access.  Evaluation protocols are still fragmented. 
For example, MatSci-NLP \citep{matscinlp2023} only covers text-based tasks, and there are no common benchmarks for multimodal reasoning or cross-modal generalization.  
TDCM25 \citep{polat2025tdcm} and MaCBench \citep{macbench2024} are two other efforts that show progress toward shared resources.  
However, there is currently no consensus in the sector about evaluation suites.  
Future research directions encompass the integration of physical invariances into fusion architectures and the establishment of standardized multimodal datasets featuring interoperable ontologies, both of which are essential for ensuring that multimodal AI significantly advances scientific discovery and can be evaluated by its contribution to end-to-end discovery outcomes (e.g., experiment selection, validation throughput, and constraint satisfaction), not only predictive metrics.

\subsection{Explainability \label{sec:3_6_reactive_explain}}
Explainable AI is becoming a key component of AI-based applications in materials science.
This change is happening because the field increasingly requires models that go beyond forecasting to provide mechanistic insight, error diagnosis, and trustworthy decision-making~\citep{XAI4matsci_survey}.
This is considerably more important if we use autonomous agentic LLMs in real-world tests, where explanations can serve as verification artifacts that gate high-stakes experimental actions.
There are no universally accepted criteria for evaluating interpretations among groups. 
\citet{AlvarezMelis2018_SENN} came up with three dimensions: explicitness, faithfulness, and stability. Explicitness examines the clarity and comprehensibility of the explanations, faithfulness assesses whether the relevance score accurately reflects the true determinants of the model output, and stability assesses the consistency of explanations derived from similar inputs. 
Other research also makes a distinction between plausibility and faithfulness. 
Plausibility refers to consistency with human thinking, while faithfulness refers to adhering to the internal thought process of the model~\citep{agarwal2024faithfulnessvsplausibilityunreliability}. 
In a comparable framework, \citet{Lipton2018_Myths} delineated three tiers of model interpretability: simulatability (the entirety of the model is comprehensible to humans), decomposability (the individual components are interpretable), and algorithmic transparency (the learning process is comprehensible and reproducible).

In this survey, we consolidate these perspectives into three domain-relevant axes for evaluating explainability in AI for materials science:
\begin{itemize}
    \item \textbf{Physical validity}: Do the explanations align with established scientific reasoning and physical laws, making them coherent to human experts? 
    \item \textbf{Faithfulness}: Do the explanations accurately reflect the internal thinking process of the model, rather than being a post-hoc approximation? 
    \item \textbf{Stability}: Do similar or adjacent inputs produce consistent explanations, ensuring robustness to minor perturbations or noise? 
\end{itemize}
These axes provide a unified evaluation framework covering sparse closed-form models, attention and graph explainers, as well as physics-informed methods. They will guide our subsequent discussion of interpretability methods.

\subsubsection{Sparse and Closed-form Models\label{sec:3_6_1_sparse_model}}
In materials science, few explainable models are more prominent than those that generate sparse closed-form representations between input features and target properties. The SISSO family of algorithms\citep{Ouyang2018_SISSO, Ouyang2022_iSISSO, TorchSISSO2024}, symbolic regression methods applied for perovskite catalysis \citep{SR_Perovskites2020} scientific machine learning with sparse regression for stability equations \citep{SciML_SR2022}, and dimension-cons-trained symbolic regression for band gap prediction \citep{SR_DSC2022}. Unlike post-hoc explainers, these methods directly return human-readable algebraic formulas that may be considered candidate scientific laws or descriptors. This makes them particularly appealing in domains like materials science, where physical interpretability and mechanism discovery are paramount. 
However, these methods typically operate on curated feature sets and may struggle with raw, high-dimensional modalities, motivating complementary neural explainers discussed next.
 
Sparse and closed-form models offer two advantages. First, they frequently experience high levels of faithfulness, because the symbolic expression itself is not a proxy, but a predictive model. Second, their emphasis on physically meaningful operators (e.g., ionic radii ratios, electronegativity differences) improves the plausibility and physical validity. Yet the shortcomings of these approaches include poor stability: tiny changes in input data or the feature pool may produce different symbolic forms, which raises questions about reproducibility and robustness across datasets. The more recent algorithmic progress achieved, e.g., i-SISSO (mutual-information prescreening) and VS-SISSO (iterative variable selection), partially mitigates these problems by improving search efficiency and descriptor stability. 
Sparse and closed-form techniques are the best on physical validity and faithfulness overall, but their Achilles heel is stability, particularly under noisy or heterogeneous data. 
This limitation is especially consequential in closed-loop discovery, where unstable descriptors can misguide experiment selection and slow down exploration.
This limitation motivates complementary directions such as attention-based explainers and physics-informed neural networks, which we discuss in the following subsections.

\subsubsection{Attention and Graph Explainers\label{sec:3_6_2_attention_graph}}

Whereas sparse regression methods provide intrinsic interpretability by expressing structure-property relations in closed-form descriptors\citep{Ouyang2018_SISSO}, most state-of-the-art AI models in materials science are deep neural networks, in particular GNNs and transformer-based architectures\citep{XAI4matsci_survey, Xia2024XElemNet}. These models achieve strong predictive performance across a wide range of tasks but are often criticized as “black boxes”, since their internal decision logic is difficult to inspect directly. In response, recent work has introduced both embedded and post-hoc explanation strategies to make such models more transparent by highlighting which atoms, bonds, compositions, or spectral features drive individual predictions\citep{XAI4matsci_survey, Xia2024XElemNet}.

Representative examples include CrabNet~\citep{Wang2021_CrabNet}, a compositionally constrained attention network whose attention mechanisms surface influential elemental constituents of a compound; GANN~\citep{shi2025graph}, a graph attention networks for 2D materials property prediction that localizes importance to specific atoms and bonds; CrysXPP~\citep{Das2022_CrysXPP}, which trains crystal GNNs under property-sensitive explanatory paths to connect predictions with interpretable substructures; and spectro-scopy-focused interpreters that generate attribution maps of spectra to identify peaks that contribute most to the output~\citep{Kotobi2024_XAS}. 
Together, these approaches move beyond raw accuracy toward more transparent materials models that can be interrogated by domain experts, even though important challenges remain in ensuring that the resulting explanations are faithful, stable, and aligned with underlying physical mechanisms.

Attention-based and graph-based explainers have strengths in their capacity to provide local, instance-particular reasons, like determining which atoms, bonds or spectral peaks guided the most prediction\citep{Xia2024XElemNet, agarwal2023evaluating}. This increases the physical validity of their measurements, as highlighted elements tend to conform to established chemical intuition. Nevertheless, faithfulness is not always guaranteed: attention weights may not correspond to the model’s actual decision logic, raising the question of whether highlighted features are causally responsible or only correlated\citep{Jain2019_AttentionNotExplanation}.

Stability still remains a hurdle, because inputs perturbed in small amounts or random seeds can modify the highlighted substructures or spectra\citep{agarwal2023evaluating}. There are multiple approaches, including integrated gradients\citep{Sundararajan2017IG}. It appears that the durability of these explanations in materials models has been studied, and the proposed robustness is explored by perturbation-based tests\citep{agarwal2023evaluating}. Thus, overall, attention- and graph-based explanation approaches are highly desirable for supplying domain-plausible, instance-level insights. They are great for alignment with human chemical reasoning (plausibility/validity) but suffer from challenges in faithfulness and stability. These limitations motivate hybrid approaches that have a combination approach: attention using perturbation tests or physics-informed constraints, leading into the next subsection on physics-informed interpretability.

\subsubsection{Physics-informed Interpretability\label{sec:3_6_3_physics}}

A third group of solutions provides intrinsic interpretability by embedding known physical principles directly into the model design.  
Physics-informed methods constrain predictions to physically consistent families, which builds trust and usefulness in materials research\citep{XAI4matsci_survey}. They don't rely on explanations that come after the fact.  Symmetry-equivariant networks, constraint-based architectures, or the strict implementation of conservation and dimensional norms can all help with this\citep{Batzner2022_E3NN}.
For example, equivariant graph neural networks that take into account the translational, rotational, and permutational symmetries of crystalline systems \citep{Batzner2022_E3NN}; models that are limited by physical invariants like energy conservation or stress–strain relationships \citep{martonova2025generalized}.
% ; and spectral-prediction networks that include sum rules and causality conditions in their outputs \citep{Xie2024_Spectra}. 

The fundamental benefit of physics-informed interpretability is that it focuses on physical validity: predictions follow the known structure of the issue space without making up correlations~\citep{zhang2022analyses}.  
Physics-informed interpretability enhances faithfulness, stability, and safety by confining models to physically valid actions, acting as critical feasibility guards in closed-loop discovery. 
But these limits can make it hard to be flexible and generalize when the laws of physics are complicated or not fully understood. This means that we need hybrid approaches that combine clear physical rules with data-driven learning to make strong, agentic pipelines.

\subsection{Pipeline-Centric Perspective}

% \begin{figure}[ht]
%     \centering
%     \includegraphics[width=0.9\textwidth]{fig/section_3.7.png}
%     \caption{Moving beyond proxy benchmarks in AI for materials science. The diagram highlights the limitations of current reactive models, which are often optimized for isolated tasks (proxies) like property prediction rather than real-world discovery. It illustrates the necessity of embedding these fine-tuned components into a broader agentic loop, where they serve as adaptive tools driven by end-to-end experimental outcomes rather than static dataset scores.
%     }
%     \label{fig:sec3_perspective}
% \end{figure}

After covering Q2, we discuss the interaction between the materials science fine-tuned LLMs and the agentic LLMs as shown in Fig.~\ref{fig:overall}.  
On the one hand, existing pre-trained LLMs lack plasticity~\citep{Dohare2024,Springer2025}, focusing on how these two components jointly implement decision-making in our pipeline-centric framework
Moreover, compared to the general pre-training data, MatSci domain-specific data, as discussed in Secs.~\ref{sec:3_prediction}, \ref{sec:3_2_mining}, and \ref{sec:3_3_generation}, is orders of magnitude smaller even if we apply techniques like data augmentation as discussed in Sec.~\ref{sec:data_augmentation}. 
These observations suggest that most behaviors of fine-tuned models are largely inherited from the pre-training stage, making it difficult to reliably elicit the desired behaviors using only limited MatSci-specific fine-tuning.

Stepping back across the previous subsections, it becomes evident that most current AI4Mat-Sci formulations are defined around narrowly scoped, mostly supervised tasks such as property prediction, information extraction, structure generation, and local optimization, each tied to bespoke datasets that are limited in both quantity and curation quality compared to the web-scale corpora used for general LLM pre-training as discussed in Sec. \ref{sec:2_progress_of_AI}, and are typically optimized in isolation from end-to-end discovery outcomes.
Even when augmented by multi-fidelity modeling, synthetic data, or automated extraction pipelines, these datasets remain sparse and noisy, making them at best partial proxies for the rich, open-ended discovery problem that materials science ultimately cares about, leaving a substantial gap between benchmark success and real experimental impact.
A straightforward response would be to scale up the quality, diversity and quantity of supervised MatSci tasks and benchmarks.
However, from an end-to-end, pipeline-centric perspective that explicitly targets the discovery of novel and useful materials under finite resources, over-optimizing models for performance on such intermediate tasks risks locking effort into local improvements on fragile surrogates (e.g., scores on a particular property dataset) while neglecting whether these gains actually translate into faster, safer identification and validation of new functional materials. 
In this view, the MatSci-tuned tasks surveyed in this section should be regarded as adjustable operators within a larger discovery pipeline, not as endpoints; their design, data, and evaluation criteria ought to be continuously re-examined through the lens of the ultimate discovery reward, so that limited high-quality MatSci data is allocated to those parts of the pipeline that most directly advance real experimental outcomes rather than merely inflating intermediate scores. 
Consequently, we argue that effort should gradually shift from reactive intermediate MatSci benchmarks, as discussed in this section, toward more agentic systems that explicitly close the loop with simulation and experiment, which will be developed in detail next.

On the other hand, increasing plasticity of the pre-training or adaptation stage carries the risk of eroding fundamental safety and security properties. 
Materials-science-specific tasks cannot reasonably be expected to cover the full spectrum of normative constraints encoded in general-purpose pre-training corpora, so aggressive adaptation must be balanced against the need to preserve these foundational principles. 
From the pipeline-centric standpoint, the challenge is therefore to couple relatively scarce, noisy MatSci supervision with general-purpose pre-training in a way that steers the overall system toward the end-to-end objective of autonomous, safe materials discovery, without sacrificing the broad alignment and robustness inherited from the foundation models.

\section{Agentic Systems for Materials Science\label{sec:4_agent_overall}}

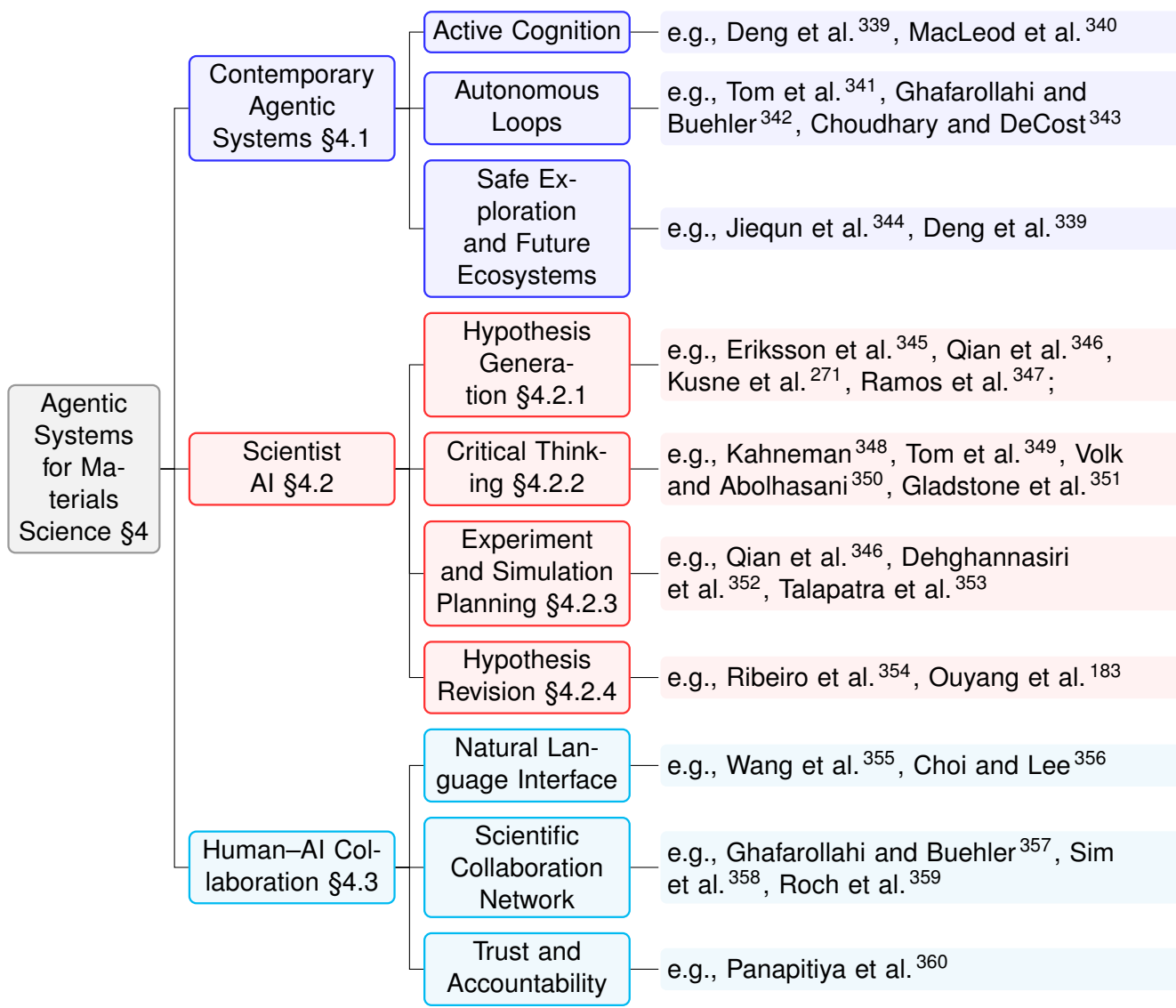
\begin{figure}
    \footnotesize
    \begin{forest}
        for tree={
            forked edges,
            grow'=0,
            draw,
            rounded corners,
            node options={align=center,},
            text width=2.7cm,
            s sep=6pt,
            calign=child edge, calign child=(n_children()+1)/2,
            font=\small,
        },
        [Agentic Systems for Materials Science~\S\ref{sec:4_agent_overall}, fill=gray!45, parent
            [Contemporary Agentic Systems~\S\ref{sec:agentic_sys_in_mat_discovery}, for tree={ pretrain}
                [Active Cognition, for tree={ pretrain}
                    [
                        {e.g., \citet{Deng2023}, \citet{MacLeod2020}}, pretrain_work
                    ]
                ]
                [Autonomous Loops, for tree={ pretrain}
                    [{e.g., \citet{tom2024}, \citet{Ghafarollahi2024}, \citet{Choudhary2021}}
                    , pretrain_work]
                ]
                [Safe Exploration and Future Ecosystems, for tree={ pretrain}
                [{e.g., \citet{Zhang2018}, \citet{Deng2023}}
                , pretrain_work]
            ]
            ]
            [Scientist AI~\S\ref{sec:scientist_ai_overview}, for tree={fill=red!45,template}
                % [General~\S\ref{sec:experiment_planning},  template
                %     [{e.g., \citet{Wang2023NatureAIDiscovery}, \citet{Lu2024AIScientist}, \citet{King2009AutomationScience}}
                %     , template_work]
                % ]
                [Hypothesis Generation~\S\ref{sec:hypothesis_generation},  template
                    [{e.g., \citet{Eriksson2019NeurIPS}, \citet{Qian2023Patterns}, \citet{Kusne2020NatCommun}, \citet{Ramos2025ChemSci}};
                    , template_work]
                ],
                [Critical Thinking~\S\ref{sec:critical_thinking},  template
                    [{e.g., \citet{Kahneman2011ThinkingFastSlow}, \citet{Tom2024SelfDrivingLabs}, \citet{Volk2024Metrics}, \citet{EBT2026}}
                    , template_work]
                ]
                [Experiment and Simulation Planning~\S\ref{sec:experiment_planning},  template
                    [{e.g., \citet{Qian2023Patterns}, \citet{Dehghannasiri2017CommMatSci}, \citet{Talapatra2018PhysRevMat}}
                    , template_work]
                ]
                [Hypothesis Revision~\S\ref{sec:interpretation_revision},  template
                    [{e.g., \citet{Ribeiro2016KDDLIME}, \citet{Ouyang2018_SISSO}}
                    , template_work]
                ]
            ]
            [Human–AI Collaboration~\S\ref{sec:4_3}, for tree={fill=blue!45, answer}
                [Natural Language Interface, answer
                    [{e.g.,
                    \citet{wang2024matgpt},
                    \citet{Choi2024}
                    }
                    , answer_work]
                ]
                [Scientific Collaboration Network, answer
                    [{ e.g.,                    \citet{ghafarollahi2024sciagentsautomatingscientificdiscovery},
                    \citet{sim2024chemos},
                    \citet{roch2020chemos}
                    }
                    , answer_work]
                ]
                [Trust and Accountability, answer
                    [{e.g.,
                    \citet{panapitiya2025autolabscognitivemultiagentsystems}
                    }
                    , answer_work]
                ]
            ]
        ]
    \end{forest}
    \caption{Taxonomy of recent progress of agentic systems for materials science.}
    \label{tree:secion-4}
    \end{figure}

In this part, we try to address the guiding question \textbf{Q3} in Sec. \ref{sec:intro}: What capacities have been developed, and how can we bridge the gap between existing AI for materials science and a fully end-to-end autonomous materials discovery pipeline?

In the previous sections, we reviewed general AI (Sec. \ref{sec:2_progress_of_AI}) and AI for materials science (Sec. \ref{sec:3_AI_for_matsci}), the latter of which focuses on reactive, isolated tasks such as property prediction, extraction, and generation.
While these specialized modules have achieved high performance on static benchmarks, they remain fragmented components that do not, on their own, constitute a discovery engine. 
From our proposed end-to-end, pipeline-centric perspective, true acceleration in discovery requires integrating these capabilities into coherent, goal-directed loops that can plan, execute, and learn from experiments. 
Accordingly, we now review the last advances in agentic systems for materials science, examining how they transition from tools that passively handle assigned tasks to active processors capable of closing the discovery loop.

\subsection{Contemporary Agentic Systems in Materials Discovery\label{sec:agentic_sys_in_mat_discovery}}

\paragraph{From Passive Prediction to Active Cognition\label{sec:sec_4_1_1}}
Over the past ten years, AI's role in materials science has changed from being a passive tool for analysis to an active participant in discovery.
Early research mainly focused on prediction with graph neural networks (e.g., CGCNN \citep{Xie2018}, ALIGNN \citep{Choudhary2021}) and pre-trained models (e.g., CHGNet \citep{Deng2023}). 
They can learn the mappings between structures and properties from static datasets. 
These approaches showed that complex atomic relationships in materials could be encoded into high-dimensional latent representations. However, these AI models still operated passively. 
They could only predict but not interact. 
Therefore, agentic materials discovery establishes a new paradigm: AI is redefined as an autonomous scientific partner that capable of independently generating hypotheses, planning experiments, and continuously learning through feedback loops\cite{Lu2024AIScientist, boiko2023autonomous}.

An agentic system in materials research operates according to the principle of closed-loop discovery. 
Formally, let a material's state be represented by a descriptor vector $x \in \mathbb{R}^n$, with measurable properties $y = f(x)$ obtained through experiment or simulation. Traditional supervised learning seeks to approximate $f$. 
In contrast, an agentic system defines a control policy $\pi_\theta(a|s)$ that selects actions $a$ (e.g., synthesis parameters, composition proposals) based on system state $s$ (e.g., prior experimental results, model uncertainty) to maximize a long-term objective $R = \mathbb{E}[\sum_t \gamma^t r_t]$. 
In materials settings, this reward typically balances experimental performance, resource cost, and safety considerations rather than only predictive accuracy.
RL thus becomes a natural formalism for autonomous experimentation \citep{beeler2023chemgymrlinteractiveframeworkreinforcement}. 
However, in practice, the state and action spaces are highly non-convex, expensive to explore, and embedded with scientific constraints such as thermodynamic stability or synthesis feasibility. Consequently, the success of agentic systems depends on the integration of uncertainty quantification, active learning, and surrogate modeling to guide exploration efficiently\cite{montoya2020autonomous, allec2025active}.
This integration distinguishes fully agentic pipelines~\citep{ChemRxiv2025AgenticAssistant} from simpler Bayesian optimization loops by coupling rich state representations.

\paragraph{Autonomous Loops: Self-Driving Labs and Multi-Agent Hierarchies\label{sec:sec_4_1_2}}
Self-driving laboratories (SDLs) play an important role in autonomous loops. An SDL combines robotic hardware for synthesis and characterization with AI models that predict outcomes, evaluate uncertainty, and iteratively plan the next experiment \citep{Hase2021}. In these systems, the learning algorithm closes the ``materials loop'': (1) Design and selection: propose an experiment or select programs in the database, (2) Execution: perform synthesis autonomously, (3) Evaluation: measure outcomes, and (4) Evolution: retrain the model on updated data. 
This cycle implements concrete credit assignment from measured outcomes back to the models that propose and prioritize experiments.
Studies in autonomous thin-film optimization \citep{Nikolaev2016} and catalyst discovery \citep{tom2024} have demonstrated that such loops can simplify experimental search spaces and preserve physical interpretability at the same time. AI's job in autonomous loops shifts from passive reasoning to agentic cognition with the ability to autonomously generate and evaluate hypotheses.

Recently, hierarchical and multi-agent architectures have been developed to extend this autonomy recently. 
For example, MatExpert~\citep{ding2025matexpert} utilized multiply stages to mimic the human materials design expert's workflow. In multi-agent system design, there is no longer just one single decision-making entity. The discovery process is decomposed into several specialized agents, such as \emph{Predictor}, \emph{Planner}, and \emph{Verifier}. Those agents can communicate with each other via standardized protocols \citep{Ghafarollahi2024}. The \emph{Predictor} utilizes neural networks (e.g., CHGNet \citep{Deng2023}, ALIGNN \citep{Choudhary2021}) to evaluate candidate materials, while the \emph{Planner} uses Bayesian optimization or RL to choose the promising directions in the action space. The \emph{Verifiers} are embodied by robotic automation and are capable of conducting experiments and sending the results back to the shared memory. These architectures enable modular expansion and enhance explainability. Each agent follows a well-defined scientific rubric and simulates the division of human labor within research teams.

\paragraph{Safe Exploration and Future Ecosystems\label{sec:sec_4_1_3}}
Agentic systems must also balance exploration and exploitation. 
Pure exploitation may converge to local optima because it continuously refines known successful materials, while unbounded exploration wastes too many resources on random new paths. 
Techniques such as upper confidence bound (UCB) sampling and Thompson sampling in Bayesian optimization frameworks \citep{Snoek2021} provide theoretical guarantees for efficient exploration. Meanwhile, recent RL approaches integrate epistemic uncertainty from ensemble models \citep{Lakshminarayanan2017} to quantify the expected information gain of each new experiment. In materials discovery, this translates into a quantifiable trade-off between novel compositions and reliable performance. Such uncertainty-aware strategies embody the concept of “safe exploration” in the physical sciences\cite{bensberg2024uncertainty}.

Another important development is the emergence of simulation and experiment in hybrid simulation-experiment loops\cite{joshi2021artificialintelligencebasedautonomous}. 
High-fidelity simulations, including density functional theory (DFT) or molecular dynamics (MD) can act as virtual laboratories, generating synthetic data to pre-train models or provide experimental priors. 
Frameworks like DeepMD \citep{Zhang2018} and CHGNet~\citep{Deng2023} provide fast computation that makes it capable of running millions of virtual experiments before a single synthesis is attempted. Agentic systems that combine simulation with experiment bridge the gap between computation and reality. 
Virtual agents can co-evolve with their physical counterparts, which is closely related to the concept of digital twins for materials discovery~\citep{kalidindi2022digital}.

The future trajectory of agentic systems will likely involve scientific multi-agent ecosystems\cite{su2025headsbetteroneimproved}, in which autonomous entities collaborate, compete, and refine hypotheses and ideas under shared goals. By combining LLMs for reasoning and documentation with RL planners and simulation engines, these ecosystems could perform discovery, hypothesis explanation, literature grounding, and experimental justification in natural language\cite{vasu2025harpatestabilitydrivenliteraturegroundedframework, alkan2025surveyhypothesisgenerationscientific, treloar2022deep}. 
Such ``AI laboratories'' will require governance mechanisms\citep{Shneiderman2020} to ensure safety and reproducibility, but they are also capable of compressing the scientific feedback cycle from months to hours.

In summary, agentic systems represent the emergence of AI, robotics, and scientific reasoning into a unified framework for autonomous scientific discovery. 
This trend shows a change from using models only to speed up predictions to using them as cognitive and experimental co-pilots that actively guide the discovery process.
In this process, machines not only compute results but also design the path to discovery. The next subsection explores how this transition naturally leads to the concept of Scientist AI: systems capable of generating and testing scientific hypotheses beyond pattern recognition towards end-to-end discovery.

\subsection{Scientist AI: Beyond Data Fitting to Scientific Reasoning}
\label{sec:scientist_ai_overview} 
 The new phase in AI for materials science goes from making accurate predictions to using AI to do scientific reasoning on its own. 
 This means systematically acquiring, testing, and changing beliefs about the physical world\cite{Tom2024SelfDrivingLabs}.  
 Early deep learning frameworks such as CGCNN~\citep{Xie2018} and ALIGNN~\citep{Choudhary2021}, discussed in Sec.~\ref{sec:agentic_sys_in_mat_discovery}, are capable of learning intricate structure–property mappings. However, they cannot articulate hypotheses, mechanisms, or explanatory frameworks\cite{Tom2024SelfDrivingLabs}.

 In contrast, a scientist AI system \citep{Wang2023NatureAIDiscovery} aims to operationalize scientific reasoning. 
 It independently formulates hypotheses, orchestrates virtual or tangible experiments, and updates its internal models in response to the resultant evidence~\citep{Lu2024AIScientist}.Instead of just guessing what attributes would happen with certain inputs, this kind of system keeps and improves a list of possible explanations for how structure, processing, and environment affect behavior\cite{Tom2024SelfDrivingLabs}. It also looks for data that will help test and improve those explanations\cite{Tom2024SelfDrivingLabs}.  Previous research on robot scientists~\citep{King2009AutomationScience} and self-driving laboratories~\citep{Tom2024ChemRevSDL} illustrates aspects of this concept, wherein automated systems repetitively suggest experiments, gather data, and revise internal models accordingly. 
 This sub-section delineates emerging methodologies—active learning, autonomous experimentation, symbolic inference, and literature-grounded reasoning—as approximating various elements of a reasoning-centric workflow.  
 Nonetheless, these elements have yet to be amalgamated into a cohesive scientific methodology: hypotheses are infrequently articulated, causal or mechanical reasoning is limited, and the integration of priors across modalities remains fragile.

 \subsubsection{Hypothesis Generation and Search Over Scientific Spaces}
 \label{sec:hypothesis_generation}

One of Scientist AI's main skills is using priors, evidence, and uncertainty to recommend pathways in the materials space\citep{Tom2024SelfDrivingLabs}. Predictive models can inform practitioners what the attributes of a target will be based on certain inputs. However, making hypotheses demands working in organized scientific domains where composition, processing parameters, mechanisms, and physical restrictions all matter\citep{Tom2024SelfDrivingLabs}. Recent advancements in machine learning and autonomous testing exhibit significant potential for proposal generation; yet, they still lag far behind human scientific thinking\citep{Volk2024Metrics}.

Contemporary Bayesian optimization (BO) and active learning frameworks have exhibited robustness in suggesting useful experiments characterized by quantified uncertainty.  
These methods usually use Gaussian process surrogates or similar probabilistic models~\citep{Eriksson2019NeurIPS}. The goal is to find a balance between exploration and exploitation in high-dimensional design spaces~\citep{Qian2023Patterns}.  In materials science, BO has facilitated the swift optimization of alloys, catalysts, and thin-film systems\citep{Volk2024Metrics}. 
It helps reduce the number of trials by several orders of magnitude and allowing for closed-loop experimental workflows~\citep{Kusne2020NatCommun}.  
These systems partially implement the hypothesis-proposal phase of scientific reasoning by identifying promising areas inside the design space amongst existing uncertainty\citep{Chen2025ActiveLearning}.

Generative models offer an alternative approach to hypothesis formulation.  
Variational autoencoders, generative adversarial networks, and diffusion-based models have been utilized in molecular and crystalline design~\citep{Anstine2023JACS}, yielding candidate structures tailored to certain target properties or design goals~\citep{Nigam2021ChemSci}.  
These methods increase the effective hypothesis space, allowing for the investigation of chemically valid yet previously unrecognized compositions and microstructures.  
Recent research illustrate inverse-design workflows wherein generative models provide candidates that fulfill various objectives or constraints, including stability and performance thresholds, succeeded by evaluation using high-fidelity simulations or experiments~\citep{Anstine2023JACS}.

LLMs add another way to come up with hypotheses. 
LLMs and science-specific variations, like SciBERT, can read a lot of scientific texts and make sense of them\citep{beltagy-etal-2019-scibert}. 
They can also find domain-specific entities and relationships and provide mechanistic narratives or qualitative design rationales\citep{Dagdelen2024StructExtract}.  
Recent agentic systems that link LLMs to chemistry and materials tools demonstrate that these models can suggest reaction pathways, enumerate viable synthetic routes, and advocate for experimental parameter sweeps by utilizing patterns found in the literature and outputs from external tools\citep{Boiko2023Autochem}.  
These systems are increasingly including explicit priors and limitations when used with retrieval and organized knowledge bases\citep{Dagdelen2024StructExtract}. 
This makes it easier to come up with hypotheses that are based on current materials knowledge and makes it easier to integrate with that knowledge~\citep{Ramos2025ChemSci}.

Even with these improvements, today's systems still do not use hypothesis generation in a scientifically sound way.  
First, hypotheses are seldom articulated openly; the majority of approaches provide candidate points or structures without delineating the corresponding mechanical assumptions, validity regimes, or expected failure mechanisms\citep{Volk2024Metrics}.  
Second, generative proposals are frequently assessed solely via surrogate property predictors, exhibiting minimal linkage to falsifiable assertions or established theories, hence limiting their function in systematic hypothesis testing and refinement\citep{Volk2024Metrics}. 
Third, constructing long-horizon hypotheses—like multi-step reasoning from mechanism to candidate design to experimental protocol—continues to be fragile and heavily reliant on limited training distributions or manually generated prompts\citep{TrustworthyAI2025}.  
Lastly, even though LLM-based agents seem to be able to combine literature and coordinate tools, their factual reliability, calibration, and grounding in domain knowledge are still problems that need to be addressed. This is shown in broader evaluations of foundation models~\citep{Bommasani2021FMs} and domain-specific LLMs for chemistry~\citep{Ramos2025ChemSci}.

All of these changes show that AI systems may suggest candidate materials, designs, and mechanistic theories based on data and with an awareness of uncertainty\citep{Tom2024SelfDrivingLabs}.  Nonetheless, they have not yet produced organized, falsifiable hypotheses or engaged in systematic reasoning on alternative explanatory theories\citep{TrustworthyAI2025}.  
To improve this skill, generative modeling has to be more closely linked to physical limitations, clear hypothesis representations, and systematic uncertainty quantification within a larger agentic loop for autonomous materials discovery\citep{Tom2024SelfDrivingLabs}.

\subsubsection{Scientific Critical Thinking\label{sec:critical_thinking}}

Unlike general-domain LLM reasoning, chemical reasoning must deal with uncertainty, confounded and scarce datasets, and slow, costly experimental feedback\citep{Tom2024SelfDrivingLabs, Volk2024Metrics}. 
Literature data is often noisy or conflicting, and learning frequently depends on imperfect proxy rewards from simulations, since reliable signals from laboratory experiments are scarce and costly\citep{Tom2024SelfDrivingLabs}. 
At the same time, molecular discovery is inherently multi-objective, where improving one function  can come at the cost of damaging another\citep{Jiang2025GenerativeAL}.  

Human cognition is often described as two modes of thought: System 1 (“fast thinking”) and System 2 (“slow thinking”) \citep{Kahneman2011ThinkingFastSlow}. 
System 2 is crucial for complex problems that require deeper and deliberate decision-making. 
Recent work on Energy‑Based Transformers frames “thinking” as an optimization that not only feeds richer context but also assesses the plausibility of its own predictions and dynamically determines when to halt reasoning\citep{EBT2026}. 
Future research should aim to develop a System 2–style thinking agent to anticipate the functions and interactions of generated molecules in future physical states and encode these imagined outcomes as additional input, enabling deliberate reasoning and hypothesis generation \citep{EBT2026}. 
The model should also generate and evaluate alternative hypotheses, fill in missing contextual information, and connect disparate findings\citep{Baatar2025LLMScientificMethod}. 
Rather than trusting all sources equally, it should assign weights to claims, hypothesize negative results, and actively resolve conflicting evidence\citep{Baatar2025LLMScientificMethod}. 
To improve its causal understanding, the model should simulate counterfactual alternative molecular designs and their expected outcomes\citep{Sevilla2024MCD}, prioritize underexplored regions of chemical space through active learning, especially those characterized by conflicting evidence or high uncertainty in counterfactual simulations\citep{Jiang2025GenerativeAL}.

\subsubsection{Experiment and Simulation Planning (Decision-Making Under Uncertainty)}
 \label{sec:experiment_planning}

Another important skill for a Scientist AI is being able to plan experiments and simulations when things are uncertain\citep{Tom2024SelfDrivingLabs, Volk2024Metrics}.  
In this situation, planning is choosing the next actions—new measurements, simulations, or adjustments to the process—that are thought to be the most useful or informative based on the present hypotheses, available data, and priors\citep{Dehghannasiri2017CommMatSci}.  
Experiment and simulation planning, on the other hand, needs decisions to be made in order with restricted budgets, different costs, and safety or feasibility limits on which activities can be taken\citep{Talapatra2018PhysRevMat}.

Modern materials informatics has created a wide range of tools for flexible experimental design\citep{Tom2024SelfDrivingLabs}.  
Active learning and Bayesian optimization frameworks utilize uncertainty-aware surrogate models and acquisition functions to dynamically select fresh measures that either optimize information gain or enhance a specified property \citep{Volk2024Metrics}.  
Optimal experimental design formalizes the challenge as a trade-off between elucidating the response surface and differentiating among competing models, frequently within the confines of limited resources\citep{Dehghannasiri2017CommMatSci, Talapatra2018PhysRevMat}.  
These methods already allow practitioners to decide which composition, processing condition, or microstructure to evaluate next based on a current surrogate and its uncertainty estimates.  
However, they typically optimize scalar objectives or predictive uncertainty rather than explicitly targeting hypothesis discrimination or mechanistic insight\citep{Volk2024Metrics}.

In self-driving laboratory (SDL) architectures, planning experiments and automating the lab have been closely linked.  
Closed-loop platforms for thin-film and semiconductor discovery combine robotic synthesis and characterization with BO-based planners to autonomously search process spaces and improve multi-step recipes\citep{Kusne2020NatCommun}.  
Multi-objective and restricted Bayesian optimization has facilitated these platforms in managing trade-offs among competing performance measures while adhering to intricate experimental limitations, thus enhancing Pareto fronts for conductivity, transparency, or mechanical properties \citep{macleod2022self}.  
Recent reviews combine these changes and show that SDLs can be used as a general framework for planning closed-loop experiments in chemistry and materials science \citep{Tom2024SelfDrivingLabs}.  
From a pipeline-centric viewpoint, SDLs implement end-to-end credit assignment by directly linking model updates and planner behavior to experimental outcomes, as opposed to relying exclusively on offline proxy metrics \citep{Tom2024SelfDrivingLabs}.

Emerging LLM-based agents enhance experiment and simulation planning by integrating language models with laboratory control systems, computational tools, and data services.  
ChemCrow\citep{bran2024chemcrow} and other chemistry agents show that LLMs with special tools can design reaction sequences, call simulators, query databases, and suggest follow-up experiments based on intermediate results.  
Autonomous research frameworks utilizing general-purpose LLMs\citep{Boiko2023Autochem} have demonstrated the capacity to strategize and implement multi-step experimental campaigns, encompassing the selection of conditions, coordination of robotic platforms, and iterative enhancement of protocols \citep{Tom2024SelfDrivingLabs}.  
Recent research assesses LLM agents for the automation of particular experimental modalities, like atomic force microscopy, underscoring both the potential and the existing fragility of language-model-driven laboratory control\citep{Mandal2025AILA}.  
These systems are starting to integrate symbolic thinking, protocol generation, and tool invocation into a single planning cycle.

Even with these improvements, today's experiment and simulation planners still fall short of expert scientists in multiple dimensions.  
Most Bayesian optimization and active learning frameworks focus on optimizing scalar objectives or surrogate uncertainty; they do not explicitly represent or update structured hypotheses\citep{Dehghannasiri2017CommMatSci}.  
Acquisition functions are usually defined over design variables instead of explanatory models or mechanisms.  
Consequently, campaigns seldom aim to differentiate between rival mechanistic hypotheses, delineate phase boundaries with measurable confidence, or conduct stress tests in areas of theoretical ambiguity.  
Ad hoc filtering is typically used to deal with practical problems like scheduling, equipment failures, maintenance windows, or safety limitations, instead of using a systematic decision-theoretic formulation \citep{Tom2024SelfDrivingLabs}.  
LLM-based planners can create flexible protocols and coordinate tools, but they are still restricted by hallucinations, poor calibration of uncertainty\citep{Boiko2023Autochem}, and inadequate guarantees that they will follow safety or physical rules\citep{bran2024chemcrow}.
Achieving scientist-level experiment and simulation planning will probably necessitate the amalgamation of uncertainty-aware optimization, explicit hypothesis representations, constraint-aware scheduling, and verifiable control of automated platforms into a cohesive framework for comprehensive materials discovery.

 \subsubsection{Interpretation, Explanation, and Hypothesis Revision}
 \label{sec:interpretation_revision}

 Within the Scientist AI system, the interpretive part of the reasoning process is what pits new data against old hypotheses and comes up with new ones that better explain the facts.  Section~\ref{sec:3_6_reactive_explain} discussed how to make AI explainable in materials science in general. It introduced physical validity, faithfulness, and stability as ways to evaluate explanations~\citep{Zhong2022npjCompMatXAI}.  In this context, these axes measure how well explanations help revise hypotheses instead of just checking models: a Scientist AI must use explanations to figure out which hypotheses are still valid, which are clearly wrong, and where more data would best improve the set of hypotheses~\citep{Oviedo2022AccMaterResXAI}~\citep{Roscher2020IEEEAccessXAI}.

 Post hoc explanation techniques, such as feature attribution (e.g., LIME~\citep{Ribeiro2016KDDLIME}, SHAP~\citep{Lundberg2017NIPSSHAP}) and saliency analyses, are now commonly utilized in composition-based models, graph neural networks, and other architectures~\citep{Zhong2022npjCompMatXAI}.  In materials science, these methods show which structural, compositional, or processing features affect individual predictions and whether models seem to be based on chemically plausible patterns~\citep{Zhong2022npjCompMatXAI}~\citep{Oviedo2022AccMaterResXAI}.  Attribution methods for graph-based models that pinpoint importance to atoms, bonds, or subgraphs further link predictions to specific motifs and have been evaluated on tasks with known ground-truth functional groups or structural units~\citep{agarwal2023evaluating}.  From the standpoint of hypothesis revision, these tools facilitate a constrained form of testing: novel data can be evaluated for coherence between emphasized features and the mechanistic narrative embedded in a specific model class; however, the explanations themselves typically do not preserve or revise an explicit hypothesis set.

 As discussed in Sec. \ref{sec:3_6_reactive_explain}, sparse and closed-form models like SISSO and symbolic regression are unique in the explainability landscape~\citep{Ouyang2018_SISSO}~\citep{wang2019symbolic}.  Within the framework of Scientist AI, their primary contribution is to furnish explicit candidate hypotheses as human-readable descriptors or analytical laws that are subject to direct examination, scrutiny, and modification.  Descriptor-discovery workflows that incorporate SISSO-style algorithms or symbolic regression into active learning loops for electrocatalysis and other applications exemplify this potential: the system not only modifies parameters within a static model but can also transition between different functional forms as new data emerges~\citep{Nair2025npjCompMatSISSOAL}.  AI Feynman and other general equation-discovery frameworks show that it is possible to get governing equations from data in idealized settings\citep{Udrescu2020}. This suggests a way to make reasoning modules that work with and choose between symbolic hypotheses instead of just updating neural weights that are hard to understand.

 Even though these systems have improved, they still don't fully meet the needs of Scientist AI when it comes to interpreting and revising hypotheses. 
 Most methods discussed in Sec.~\ref{sec:3_6_reactive_explain} mainly focus on local or model-centric explainability, and their faithfulness and stability when the distribution shifts or the dataset is biased are still limited~\citep{Roscher2020IEEEAccessXAI, Zhong2022npjCompMatXAI}.  Symbolic techniques are getting closer to clear, testable hypotheses, but they still only work in low-dimensional environments and need a lot of expert input on feature spaces and operator sets~\citep{wang2019symbolic, Ouyang2018_SISSO}.  Emerging LLM- and agent-based systems can generate textual rationales, alternative mechanisms, or literature-grounded critiques in response to new data~\citep{Bran2024NatureChemCrow, Bommasani2021FoundationModels}, yet these explanations are seldom tied to a formal representation of a hypothesis set, nor evaluated on their ability to drive correct revisions of that set over multiple iterations of data collection.  
To reach scientist-level interpretation and hypothesis revision, it is therefore essential to move beyond the primarily diagnostic explainability criteria and methods of Sec. \ref{sec:3_6_reactive_explain} toward explicit hypothesis tracking, uncertainty-aware selection among competing mechanistic models, and benchmarks that directly measure how effectively an AI system revises and refines scientific hypotheses within a closed-loop discovery process.

\subsection{Human–AI Collaboration in Scientific Workflows\label{sec:4_3}}

Existing molecular AI systems remain static learners trained on historical literature data and often fail to adapt to new experimental evidence or improve through real-world feedback. 
Specifically, future work should establish a closed-loop digital–physical workflow where a high-throughput, autonomous laboratory can synthesize AI-selected molecules with minimal human involvement~\citep{bender2021artificial,tan2025ai}. 
Recent work~\cite{mCLM2025} proposed to position mCLM as a core agent of a human-in-the-loop autonomous laboratory, iterating through reasoning, proposal, synthesis, testing, and feedback from real-world synthesis and testing machines to enable never-ending improvement alongside human scientists, by leveraging LLM agents' self-improvement~\cite{wang2024enablinglanguagemodelsimplicitly,selfimprovetesttime2025} and self-evolving~\cite{selfevolve2025} capabilities. 

As artificial intelligence systems mature from tools of prediction to agents of discovery, a central question has increasingly been posed: how should humans and machines collaborate in the pursuit of scientific progress \citep{vaccaro2024combinations}? The further improvement of AI4MatSci will not be achieved only by algorithmic and computational power. It will also depend on the design of \emph{human–AI workflows} that involve human intuition and domain insight alongside machine-scale reasoning, data integration, and hypothesis generation \citep{Holzinger2019,Shneiderman2020}. 
This convergence motivates the design and use of co-creative intelligence \citep{singh2025systematicreviewhumanaicocreativity}, in which AI could serve as a scientific collaborator.

Early examples of such collaboration have already affected the materials design cycle. In prediction and generation tasks, researchers use interpretable graph neural networks (GNNs) such as ALIGNN~\citep{Choudhary2021} or CrabNet~\citep{Wang2021_CrabNet} to provide explanations for learned structure–property relationships, facilitating scientific validation and hypothesis refinement. These models support a two-way interaction between human insight and model-driven suggestions.
AI can highlight correlations that are hard to detect intuitively in high-dimensional feature spaces while human scientists can test and interpret their physical plausibility. Studies in explainable materials modeling show that transparent model architectures, such as attention mechanisms revealing atomic contributions or saliency mapping over crystal graphs, can perform scientific structural reasoning from numerical output. \citep{Holzinger2019}. Such methods are critical for enabling trust calibration, where researchers learn not to over-trusting and under-trusting~\citep{bollaert2023measuring}. 

LLMs further facilitate the interaction between AI and humans by introducing natural-language interfaces into the workflows. LLMs that are fine-tuned on domain-specific training resources such as experimental protocols and computational reports can effectively parse, summarize, and contextualize scientific results. Recent systems like MatGPT~\citep{wang2024matgpt} are capable of solving materials language processing tasks by conducting workflows, querying crystallographic databases, and helping experimental setups by giving suggestions in natural language \citep{Choi2024}. By integrating multimodal reasoning that combines textual, numerical, and structural embeddings, LLMs can also serve as adaptive documentation and knowledge retrieval systems. These systems can evolve with laboratory data streams. In this manner, human–AI collaboration is expansive, from mere task execution to the whole scientific lifecycle: from idea creation to experiment construction and execution, validation, interpretation, and publication~\citep{lu2024ai}.

 AI systems are becoming more and more autonomous, which makes  human scientists become less directly involved in low-level decision-making. This will reduce the participation of human agency and makes co-creation more difficult in some scenarios. 
 More and more AI systems like ChemOS~\citep{roch2020chemos} can enhance the planning and execution of experiments in self-driving labs. In this case, human scientists should foucs more on higher-level management, ethical oversight, supervision, and conceptual innovation\citep{Hase2021}. Interface design further contributes to more effective collaboration, which involves designing interpretable feedback loops between algorithmic reasoning and human comprehension~\citep{chen2025engagingaiinterfacedesign}. For instance, prospective interfaces can prevent confusion before it happens. 
 An explainable user interface allows users to interactively explore the data and model sensitivity so they can predict what and why the AI will do. 
 The AI decision can construct AI results as structured hypotheses, inviting human assessment to decide whether accept or deny them. These hybrid decision-making architectures align with the principles of human-in-the-loop AI, where automation and human expertise together produce more reliable outcomes than either could achieve alone \citep{Shneiderman2020}.

A second new concept at this stage is \emph{scientific collaboration networks}, which were originally developed to analyze collaborations among human researchers~\citep{Barab_si_2002} and has since broadened to include hybrid collaborations between humans and AI agents~\citep{ghafarollahi2024sciagentsautomatingscientificdiscovery}. In traditional scientific collaboration networks, nodes are scientists, and edges connect them if the scientists co-authored at least one paper jointly. Networks were constructed independently for separate fields for comparing the disciplinary patterns. By contrast, nodes in human-AI collaboration networks are specialized agents (software modules) along with human supervision. The framing implicitly favors human-agent teaming: humans define goals, agents create and refine hypotheses, and humans evaluate, execute, and validate results. This structure acts much like a human research team. Different members have different roles that work well together, and they talk to each other through shared interfaces to plan their work.

From an epistemological point of view, the collaboration of human–AI technology signifies a paradigm shift in scientific conduct. 
Rather than knowledge being created solely through human deliberation, traditional science cycle by hypothesis–experiment–analysis is bounded by human thought and speed of human communication~\citep{langley1987scientific}. 
Instead, co-creative workflows exist at the crossroads between symbolic reasoning on the one hand and data-driven discovery, turning this cycle into a process of feedback loop through AI \citep{langley1987scientific}. Transforming experimental data into organized form representations understandable to humans and machines, these systems collectively share an epistemic language that brings interpretability and reproducibility \citep{cui2024ai}. 

Collaborative frameworks also need to address issues of attribution, authorship, and bias in the future in order to address the trust and accountability issues~\citep{novelli2024accountability,oduoye2023algorithmic}. 
Designing new architectures and evaluation frameworks for trustworthy collaboration is becoming essential. AutoLabs~\citep{panapitiya2025autolabscognitivemultiagentsystems} introduces a critic agent specifically designed to double-check the safety and feasibility, while GIFTERS~\citep{amirian2025buildingtrustworthyaimaterials} tries to evaluate the AI part along dimensions such as generalizable, interpretable, fair, transparent,
explainable, robust, and stable trustworthiness principles. The scientific community is beginning to deal with questions like: Who owns AI-generated hypotheses? How can we peer review AI-based discoveries? Efforts such as transparent model reporting, tracking data sources, clear documentation, and continuous monitoring are essential for keeping the scientific integrity of AI-assisted research \citep{blau2024protecting}.Furthermore, applying fairness, transparency, and accountability principles into materials workflows is crucial for the responsible adoption of AI in both industry and academia \citep{lo2020ethical}.

Human–AI collaboration is reshaping scientific discovery, shifting AI from a purely predictive tool to an active partner. In materials science, interpretable models and large language systems bring continuous feedback between human intuition and machine reasoning, integrating speculation, simulation, and analysis into one workflow. 
As autonomous frameworks orchestrate experiments, preserving human agency in the process, open, interpretable interfaces are essential. 
New networks of human and AI agents also appear, showing this move towards co-creative science. To maintain this partnership and its importance, safety mechanisms, interpretability guarantees, and scientific alignment protocols are needed to assure that their progressively autonomous AI systems are accountable and trustworthy, and in line with human scientific objectives.

\subsection{Pipeline-Centric Perspective}

% \begin{figure}[ht]
%     \centering
%     \includegraphics[width=0.9\textwidth]{fig/section_4.4.png}
%     \caption{The essential role of human-in-the-loop in autonomous discovery. While depicting the cycle of agentic hypothesis generation and execution, this schematic emphasizes that human oversight remains a critical component of the pipeline. It illustrates how human experts do not just monitor for safety but actively co-improve the system, providing the high-level reasoning and validation necessary to align autonomous agents with meaningful scientific objectives.
%     }
%     \label{fig:sec4_perspective}
% \end{figure}

After addressing Q3, we continue our discussion on the end-to-end pipeline-centric perspective, focusing more on the role of agentic LLMs in materials discovery. 
Yann LeCun has famously argued that ``if intelligence is a cake, the bulk of the cake is unsupervised learning, the icing on the cake is supervised learning, and the cherry on the cake is reinforcement learning'' (NeurIPS 2016), thereby downplaying the centrality of RL relative to representation learning via large-scale pre-training. 
In contrast, \citet{silver2025era} argue that intelligence should fundamentally be grounded in interaction with the environment, thereby elevating the role of RL and other experience-driven approaches to a primary mechanism for building general intelligence. 
Rather than engaging directly with this broader conceptual debate, we instead focus on the concrete demands of materials science, where discovery is inherently interactive: hypotheses must be tested through simulations and (preferably) experiments, so learning through experience is less optional than it is in purely digital areas.

Given the domain-specific uniqueness outlined in the preceding subsections, these considerations collectively indicate that an agentic system constitutes the more appropriate design choice. 
However, current agentic systems reviewed in Sec. \ref{sec:4_agent_overall}—while representing significant progress—remain incomplete from an end-to-end, pipeline-centric perspective for two critical reasons.

First, a lot of current work doesn't use backward credit assignment all the way through the pipeline.  
By backward credit assignment, we mean linking the results of downstream discoveries—both successes and failures—back to choices made earlier in the process, such as corpus selection, objective design, and model adaptation. 
This way, those parts can be changed under the same end-to-end pressure.
As outlined in Sec. \ref{sec:2_progress_of_AI} and \ref{sec:3_AI_for_matsci}, training signals resulting from agentic successes or failures (such as unsuccessful synthesis attempts, inaccurate property predictions, and breached physical constraints) can theoretically be transmitted backward to prior stages, including pre-training data curation, domain adaptation objectives, and instruction-tuning criteria.  
This backward signal is necessary for the influence-driven correction discussed in Sec. \ref{sec2:pipeline-perspective}, where failures in closed-loop discovery immediately tell us which parts of the pre-training corpus should be made louder, quieter, or eliminated. 
Without this feedback loop, agentic systems stay separate from the basic representations and assumptions built into their models. This makes it harder for them to change and get better over long periods of time.

Second, most existing agentic systems work mostly in simulations or on synthetic benchmarks, not in real experimental settings. 
Simulation-based agents (e.g., those employing DFT or molecular dynamics) offer significant prototyping opportunities; however, they obscure essential domain constraints, including equipment limitations, measurement noise, synthesis variability, and the inherent irreducibility of real-world chemistry to computational models. 
As a result, agents that are only trained on simulated loops may find optimal materials that are stable in silico but impractical or too expensive to make in real life. 
Real agentic systems for discovering new materials must be closely linked to robotic labs, experimental feedback, and the messy realities of physical synthesis and characterization.
Agents can only learn useful tactics with affordable costs that close the gap between simulation and reality if they have a strong pre-trained foundation formed in the upstream stages, with experimentation providing the corrective signal that refines these priors toward robust, safety-aware autonomous discovery.

\section{Discussion \& Conclusion}

% https://arxiv.org/pdf/2508.03278
% https://arxiv.org/pdf/2506.20743

This survey synthesizes recent advances in AI for materials science through a pipeline‑centric lens that connects corpus curation, pre-training, domain adaptation, and instruction tuning to goal‑conditioned, agentic LLMs operating within open‑ended experimental environments. 
By jointly organizing the field around three guiding questions, it integrates machine‑learning and materials‑science perspectives into a single conceptual framework for autonomous materials discovery. 
The survey systematically reviews predominantly reactive tasks such as prediction, mining, generation, optimization, and verification, together with cross‑cutting issues in data, knowledge integration, multimodality, and explainability, clarifying both the strengths and limitations of current approaches. 
Building on this foundation, it characterizes emerging agentic systems and Scientist‑AI workflows that couple hypothesis generation, experiment and simulation planning, and human–AI collaboration, arguing that truly discovery‑oriented AI4MatSci requires aligning all components of the pipeline to real experimental reward signals rather than proxy benchmarks.

Revisiting Fig. \ref{fig:overall}, the agentic system operating within an open‑ended experimental environment uniquely accesses real‑world outcome signals that can be used for credit assignment at the current decision stage and retroactively propagated to upstream stages spanning pre-training, materials‑specific adaptation, and task design, as detailed in Secs. \ref{sec:2_progress_of_AI}, \ref{sec:3_AI_for_matsci}, and \ref{sec:4_agent_overall}. From this vantage point, the concluding section elaborates a coherent roadmap that operationalizes this pipeline‑centric view, integrating the insights from the preceding sections into a unified picture of how agentic LLM systems can be aligned with end‑to‑end materials discovery objectives.

\begin{figure}[ht]
    \centering
    \includegraphics[width=0.99\textwidth]{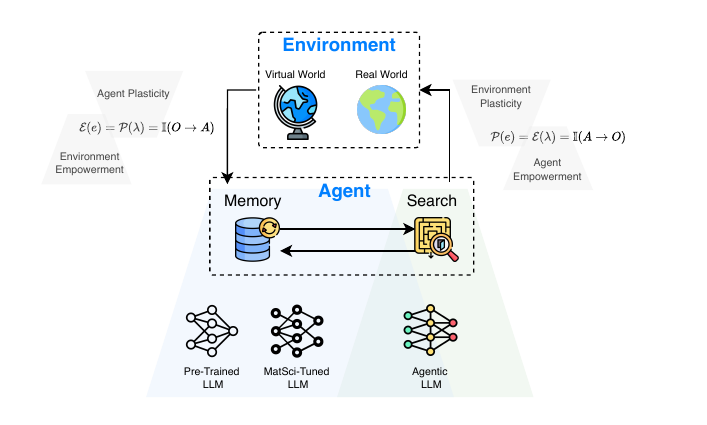}
    \caption{Search is defined as trial-and-error, generate-and-test, and variation-and-selection, whereas memory is to remember what worked best for each situation and start from there next time. In these processes, certain level of generalization is expected. The pre-trained and MatSci-tuned LLMs serve as the unified memory. The agentic LLM mostly acts as the search operator interacting with the environment, but it can access and store information during interaction into the memory. The agentic LLM can employ the search operator to interact with either the virtual world (e.g., a simulator) or the real world. Following the definition and notations in~\citep{Abel2025}, the directed information $\mathbb{I}(X \to Y)$ captures the influence that the past elements $X_{i}$ have on future elements $Y_{i}$.}
    \label{fig:search_memory}
\end{figure}

\citet{Barto2024} observed that ``reinforcement learning is memorized search'' and noted that ``in the beginning, machine learning was reinforcement learning.''
From this point of view, general pre-training, domain-specific fine-tuning and part of lifelong RL in the experimental environment serve as a unified memory, contributing to the search component of RL, as shown in Fig. \ref{fig:search_memory}. 
Note that an agentic LLM also relies on the memory function during interacting with the environment, as it stores useful information into the memory after searching. 

However, recent evaluations \citep{metr2025recentrewardhacking} show that frontier LLMs can engage in sophisticated reward hacking in benchmarking environments, indicating that the memory formed by pre-training and fine‑tuning may overfit scoring artifacts rather than reflect robust performance in real experimental settings. 
To mitigate this, experimental environment rewards should be used to reshape upstream memory, data curation, objectives, and weights, thus aligning learning with outcome‑level goals.
It has also been argued \citep{Belcak2025-df} that small language models are enough for agentic tasks, as they are ``sufficiently powerful, inherently more suitable, and necessarily more economical for many invocations in agentic systems''.  
To some extent, compressing LLMs into small LMs can be considered as a way to modify the upstream components driven by the credit assignment process mainly based on the final RL reward -- discovering novel materials in our case. 
Ideally, the whole pipeline should be tunable, driven by the final objective. 

For MatSci scenarios, we can use the influence function~\citep{Grosse2023-to} to track down the contributions of pre-training and fine-tuning data to the discovery of novel and useful materials. 
It is also promising to learn RL rules that are more suitable for MatSci settings, inspired by the success of showing how to discover a state-of-the-art RL rule that outperforms manually designed rules \citep{Oh2025-zg}. 
We encourage both the machine learning and materials science communities to view the pipeline as a whole serving the single ultimate goal -- discovering novel materials, using which to scrutinize the upstream components and redesign them through the lens of RL.

As indicated in Fig. \ref{fig:search_memory}, an agentic LLM can operate on either the virtual world and the real world. 
It is preferable and practical to first train it on the virtual world and then transfer it and continually train it in the real-world materials science environment. 
We should be aware that an agent cannot learn beyond the knowledge of a simulator, meaning that it is not realistic to expect that an agent can generalize to the knowledge outside of a simulator designed by human experts and domain data. 
This indicates the potential performance gap between an agentic LLM trained in the virtual world and the one ideally trained in the real-world environment, which is prohibitively expensive. A closely related and often underemphasized bottleneck is the limited accessibility of laboratory automation itself.
Democratizing lab automation is critical to addressing the persistent data scarcity challenge in materials science.
Despite recent progress, autonomous and self-driven laboratories remain expensive, slow to deploy, and heavily dependent on expert knowledge not only in materials science, but also in mechanical engineering, robotics, and software engineering.
As a result, adapting an existing automated laboratory to a new scientific objective is still a nontrivial and resource-intensive process, restricting the scale and diversity of real-world feedback available for learning.

We also envision parallel memory+search across the virtual and real worlds, similar to the concept of asynchronous thinking proposed by \citet{schilling2025text}. 

Current benchmarking methods for computational materials discovery mostly focus on fixed prediction tasks and isolated computational elements, overlooking the inherently iterative, exploratory, and sometimes serendipitous nature of real scientific discovery \citep{malik2025towards}.
Even though CrystalGym \citep{govindarajan2025crystalgym} provides APIs that are familiar to RL researchers, it still remains unclear if the performance can be translated into real-world experimental environments. 
Inspired by digital twins for materials \citep{kalidindi2022digital} and co-evolution of agents and the virtual environment \citep{lu2025don}, where a data interface keeps the digital and physical systems synchronized, we propose to evolve the agent, the virtual world, and all the components in the upstream stages based on the real-world environmental reward. 
While individual components and their associated views each offer limited perspectives, only a systems-level approach can deliver the complete, integrated understanding required to objectively accelerate materials innovation; this goal underpins the pipeline-centric view of our survey.
Here, a modular dynamic AI4MatSci testbed is needed, which can be updated with real-world data more easily.

In terms of additional algorithmic improvement, differently from previous digital-twins for materials discovery, we argue for a single shared agent or a population of collective agents with memory \citep{he2024ma,yu2025memagent,xu2025mem,guo2024large,tran2025multi} for both simulation and real-world experiments, and even upstream pre- and mid-training. 
This unified view is also aligned with the concept depicted in Fig. \ref{fig:search_memory}, where the whole pipeline shares a single memory across all stages, which can be updated by the final real-world reward signal. 

\citet{Bengio2025} warned about the biosecurity risks at the convergence of AI and the life sciences. 
Similarly, LLM agent-driven materials discovery and autonomous laboratory workflows can accelerate beneficial research but also lower barriers for misuse (e.g., designing materials with hazardous properties or enabling access to dangerous synthesis routes).
Thus, it is crucial to develop trustworthy and safe LLM agents for materials science research.
We depict the empowerment-plasticity view~\citep{Abel2025} in Fig. \ref{fig:search_memory}. 
Empowerment is the extent to which an agent can shape or control the set of future states it can perceive, and plasticity is the capacity of an agent to maintain adaptability in the face of significant changes or challenges~\citep{Abel2025}. 
Interestingly, the plasticity of the agent is equivalent to the empowerment of the environment, and vice-versa~\citep{Abel2025}. 
On the one hand, this implies that, for the safety concerns, the agent might need to dynamically adjust its plasticity in order to restrain the adversarial empowerment of the environment. 
On the other hand, this means that to maximize the overall yields of the whole system over time, the environment over which the agent is empowered should often have a high plasticity, e.g., with more controllable components or tools.
We argue that AI4MatSci should routinely revisit each module in the workflow and ask whether it can be brought under learning or optimization pressure rather than being treated as a frozen heuristic. 
Concretely, this means not only training neural networks but also learning data selection and curation policies~\citep{calian2025datarater}, adaptive pre‑training corpora, retrieval and tool‑calling strategies, experiment proposal policies, and even aspects of the experimental protocol itself, using the ultimate reward signals. 
Our pipeline‑centric, end‑to‑end perspective thus views the materials discovery loop as a composite system in which as many components as possible are trainable and where signals from successful or failed discoveries can, over time, reshape both models and upstream pipeline choices to better serve the ultimate goal of finding novel, useful, and safe materials.
% This empowerment-plasticity perspective also implies that memory plasticity is equivalent to search empowerment. 
Overall, our proposed unified and pipeline-centric view encourages us to design highly decoupled agents and the corresponding environments to maximize output while minimizing risks.

\bibliography{main}

\appendix
% \section{Appendix}

\end{document}